\documentclass[aps,prb,10pt,twocolumn,showpacs]{revtex4-1}

\usepackage{amsmath, amssymb}    					% needed for subequations and some math symbols
\usepackage{graphicx}   							% needed for figures
\usepackage{xcolor}     							% use if color is used in text
\usepackage{hyperref}   							% use for hypertext links, including those to external documents and URLs
\hypersetup{colorlinks=true,allcolors=blue}			% set all appearing hyperlinks to blue

\newcommand{\ddt}[0]{\frac{\partial}{\partial t}}
\renewcommand{\t}[1]{\textrm{#1}}
\newcommand{\nn}[0]{\nonumber\\}
\newcommand{\an}[0]{\allowdisplaybreaks\\}
\newcommand{\mbf}[1]{\mathbf{#1}}
\renewcommand{\k}[0]{\mathbf{k}}
\newcommand{\K}[0]{\mathbf{K}}
\newcommand{\0}[0]{\mathbf{0}}
\renewcommand{\r}[0]{\mathbf{r}}
\newcommand{\R}[0]{\mathbf{R}}

\newcommand{\q}[0]{\mathbf{q}}

\newcommand{\Jsd}[0]{J_{sd}}
\newcommand{\Jpd}[0]{J_{pd}}
\newcommand{\NMn}[0]{N_\t{Mn}}

\newcommand{\omMn}{\omega_{\t{Mn}}}
\newcommand{\ome}{\omega_{\t{e}}}
\newcommand{\omh}{\omega_{\t{h}}}
\newcommand{\gel}{g_\t{e}}

\newcommand{\gMn}{g_\t{Mn}}

\newcommand{\bs}[1]{\boldsymbol{#1}}

\newcommand{\me}{m_{\t{e}}}
\newcommand{\mh}{m_{\t{h}}}
\newcommand{\etae}{\eta_{\t{e}}}
\newcommand{\etah}{\eta_{\t{h}}}
\renewcommand{\Im}{\textrm{Im}}
\renewcommand{\Re}{\textrm{Re}}
\newcommand{\ud}{{\uparrow/\downarrow}}
\newcommand{\du}{{\downarrow/\uparrow}}

\newcommand{\Spar}[0]{\langle S^\parallel \rangle}

\begin{document}

\title{Quantum kinetic equations for the ultrafast spin dynamics of excitons in diluted magnetic semiconductor quantum wells after optical excitation}
\author{F. Ungar}
\author{M. Cygorek}
\author{V. M. Axt}
\affiliation{Theoretische Physik III, Universit\" at Bayreuth, 95440 Bayreuth, Germany}

\begin{abstract}
	Quantum kinetic equations of motion for the description of the exciton spin dynamics in II-VI diluted magnetic semiconductor quantum wells with laser driving are derived.
	The model includes the magnetic as well as the nonmagnetic carrier-impurity interaction, the Coulomb interaction, Zeeman terms, and the light-matter coupling, allowing for
	an explicit treatment of arbitrary excitation pulses.
	Based on a dynamics-controlled truncation scheme, contributions to the equations of motion up to second order in the generating laser field are taken into account.
	The correlations between the carrier and the impurity subsystems are treated within the framework of a correlation expansion.
	For vanishing magnetic field, the Markov limit of the quantum kinetic equations formulated in the exciton basis agrees with existing theories based on Fermi's golden rule.
	For narrow quantum wells excited at the $1s$ exciton resonance, numerical quantum kinetic simulations reveal pronounced deviations from the Markovian behavior.  
	In particular, the spin decays initially with approximately half the Markovian rate and a non-monotonic decay in the form of an overshoot of up to $10\,\%$ of the initial spin 
	polarization is predicted.
\end{abstract}


\maketitle

\section{Introdction}
\label{sec Introduction}

The idea behind the spintronics paradigm\cite{Dietl_A-ten, Ohno_A-window, Zutic_Spintronics, Awschalom_Challenges-for} is to combine state-of-the-art electronics based on 
carrier charge with the manipulation and control of the spin degree of freedom\cite{Awschalom_Quantum-Spintronics, Wunderlich_Spin-Hall, Chappert_The-emergence}.
Diluted magnetic semiconductors (DMS)\cite{Dietl_Dilute-ferromagnetic, Kossut_Introduction-to, Furdyna_Diluted-magnetic} present an interesting subclass of semiconductors
in this context because they can be easily combined with current semiconductor technology while at the same time providing a wide range of spin and magnetization-related effects and applications
\cite{Kossut_On-the, Bastard_Spin-flip, Cywinski_Ultrafast-demagnetization, Nawrocki_Exchange-Induced, Perakis_Femtosecond-all, DasSarma_Temperatur-dependent, Camilleri_Electron-and,
Krenn_Photoinduced-magnetization, Debus_Resonantly-enhanced, Awschalom_Spin-dynamics, Crooker_Optical-spin, Baumberg_Spin-beats}.
In DMS, a small fraction of magnetic ions, usually Manganese\cite{Furdyna_Semiconductor-and}, is introduced into a semiconductor.
While III-V compounds such as Ga$_{1-x}$Mn$_x$As are typically p-doped\cite{Dietl_Dilute-ferromagnetic} and can thus exhibit carrier-mediated ferromagnetism\cite{Ohno_GaMnAs}, 
II-VI materials such as Cd$_{1-x}$Mn$_x$Te are found to be intrinsic and paramagnetic due to the isoelectrical incorporation of the Mn impurities.

A lot of theoretical works on DMS has been devoted to the understanding of structural properties\cite{Jungwirth_Theory-of, Jungwirth_Curie-temperature, Jungwirth_Prospects-for, 
Jungwirth_Low-temperature, Jungwirth_Systematic-Study, Bouzerar_Unified-picture}.
But in many experiments, also the spin dynamics studied via optical pump-probe experiments is of interest\cite{BenCheikh_Electron-spin, Camilleri_Electron-and, 
Krenn_Photoinduced-magnetization}.
Theoretical descriptions of such experiments are less developed in the literature and are typically based on rate-equation models\cite{Bastard_Spin-flip, Cywinski_Ultrafast-demagnetization, Nawrocki_Exchange-Induced, Camilleri_Electron-and, BenCheikh_Electron-spin, Semenov_Spin-flip, Semenov_Electron-spin, Jiang_Electron-spin, Vladimirova_Dynamics-of}, coinciding with
Fermi's golden rule for vanishing magnetic field.
However, a number of experiments have provided strong evidence that these models fail to reproduce some of the pertinent characteristics of the spin dynamics in DMS. 
Most notably, experimentally observed spin-decay rates are found to be a factor of $5$ larger than the Fermi's golden rule result for spin-flip scattering of conduction band electrons 
at magnetic impurities\cite{BenCheikh_Electron-spin}.
Camilleri \textit{et al.}\cite{Camilleri_Electron-and} have argued that their optical experiments probe excitons rather than separate electrons and holes.
In this case, the effective mass entering the spin-flip rate has to be replaced by the exciton mass\cite{Bastard_Spin-flip}, offering a potential explanation for the discrepancy noted in 
Ref.~\onlinecite{BenCheikh_Electron-spin}.

On the rate-equation level, some groups have already investigated the exciton spin dynamics in DMS theoretically\cite{Tsitsishvili_Magnetic-field, Maialle_Exciton-spin-1993,
Maialle_Exciton-spin-1994, Selbmann_Coupled-free, Yanhao_Exciton-spin, Adachi_Exciton-spin}.
However, recent studies using a quantum kinetic theory for the spin relaxation of conduction band electrons in DMS revealed that correlations between the carrier and impurity 
subsystems can induce a finite memory\cite{Thurn_Quantum-kinetic, Cygorek_Carrier-impurity, Cygorek_Influence-of, Cygorek_Dependence-of} which is not captured by rate equations.
The resulting non-Markovian effects were found to be particularly pronounced for excitations close to the band edge ($\k \approx \0$)\cite{Cygorek_Non-Markovian} and become more
significant with increasing effective mass\cite{Thurn_Non-Markovian}.
These tendencies suggest that non-Markovian features are particularly relevant for excitons since, first of all, the conservation of momentum implies a vanishing center of mass momentum
($\K \approx \0$) of optically generated excitons, and second, the exciton mass is much larger than the effective mass of conduction band electrons.

In this article, we develop a microscopic quantum kinetic theory for the exciton spin dynamics in DMS that is also capable of describing non-Markovian effects by explicitly accounting for
carrier-impurity correlations.
In contrast to previous works\cite{Thurn_Quantum-kinetic} where independent electrons and holes were considered and where higher-order correlations were treated within a variant 
of Kubo's cumulant expansion\cite{Kubo_Generalized-Cumulant}, here a dynamics-controlled truncation (DCT)\cite{Axt_A-dynamics, Axt_Nonlinear-optics} is employed for the treatment
of Coulomb correlations.
This approach is especially advantageous for the description of optically-driven systems since it ensures a correct description of the dynamics up to a given order in the generating
field.
The theory derived in this paper is applicable in a wide range of different scenarios as a number of interactions are accounted for, such as the magnetic and nonmagnetic interactions between
impurities and electrons as well as holes, the Coulomb interaction responsible for the formation of excitons, Zeeman terms for electrons, holes, and impurities, as well as the 
light-matter coupling.

Moreover, we show that, in the Markov limit and for vanishing magnetic field, the quantum kinetic description coincides with the Fermi's golden rule result of 
Ref.~\onlinecite{Bastard_Spin-flip}.
Comparing numerical simulations using the quantum kinetic theory and Markovian rate equations reveals strong non-Markovian effects in the exciton spin dynamics.
In particular, the quantum kinetic calculations predict that the exciton spin initially decays with approximately half the rate obtained from Fermi's golden rule and exhibits a
nonmonotonic behavior with an overshoot of up to $10\,\%$ of the initial spin polarization.
In contrast to the situation for conduction band electrons, where nonmagnetic impurity scattering typically strongly suppresses non-Markovian features\cite{Cygorek_Influence-of}, 
here we find that, for excitons, the presence of nonmagnetic impurity scattering enhances the characteristics of non-Markovian behavior.

The article is structured as follows:
First, we discuss the individual contributions to the Hamiltonian that determines the spin dynamics of optically generated excitons in DMS quantum wells.
Next, quantum kinetic equations based on a DCT scheme are derived for reduced exciton and impurity density matrices as well as carrier-impurity correlations. 
We then derive the Markov limit of the quantum kinetic equations of motion.
Finally, we present numerical calculations and discuss the results.

\section{Theory}
\label{sec Theory}

In this section, we present the Hamiltonian that models the optical excitation and the subsequent spin evolution of excitons in II-VI DMS.
We explain the derivation of the quantum kinetic equations and, for comparison, also give the Markov limit of the equations.

\subsection{Hamiltonian}
\label{subsec Hamiltonian}

We consider an intrinsic II-VI DMS quantum well where initially no electrons are in the conduction band.
The time evolution of the system can then be described by the Hamiltonian
\begin{align}
\label{eq complete Hamiltonian}
H &= H_0^\t{e} + H_0^\t{h} + H_\t{conf} + H_\t{C} + H_\t{Z}^\t{e} + H_\t{Z}^\t{h} + H_\t{Z}^\t{Mn} + H_\t{lm} 
	\nn
	&\phantom{=\;} + H_{sd} + H_{pd} + H_\t{nm}^\t{e} + H_\t{nm}^\t{h},
\end{align}
where
\begin{align}
\label{eq H_0}
H_0^\t{e} + H_0^\t{h} &= \sum_{l \k} E^{l}_{\k} c^\dagger_{l \k} c_{l \k} + \sum_{v \k} E^{v}_{\k} d^\dagger_{v \k} d_{v \k} 
\end{align}
is the crystal Hamiltonian for electrons and holes, respectively.
Here, $c^\dagger_{l \k}$ ($c_{l \k}$) denotes the creation (annihilation) operator of an electron in the conduction band $l$ with wave vector $\k$.
Similarly, $d^\dagger_{v \k}$ ($d_{v \k}$) creates (annihilates) a hole in the valence band $v$.
The confinement potentials for electrons and holes responsible for the formation of a quantum well is denoted by $H_\t{conf}$.

As usual for the description of near band-edge excitations of semiconductors we consider the part of the Coulomb interaction conserving the number of electrons and holes, 
which corresponds to the typically dominant monopole-monopole part in a multipolar expansion\cite{Haken_Quantum-field, Axt_Nonlinear-optics, Stahl_RPA-dynamics, Huhn_Self-Consistent}.
The Coulomb interaction then reads
\begin{align}
\label{eq H_C}
H_\t{C} = \frac{1}{2}\sum_{\k \k' \q} \Big( &V_{\q} \sum_{l l'} c^\dagger_{l' \k'+\q} c^\dagger_{l \k-\q} c_{l \k} c_{l' \k'} 
	\nn
	& + V_{\q} \sum_{v v'} d^\dagger_{v' \k'+\q} d^\dagger_{v \k-\q} d_{v \k} d_{v' \k'}
	\nn
	& - 2 V_{\q} \sum_{l v} c^\dagger_{l \k'+\q} d^\dagger_{v \k-\q} d_{v \k} c_{l \k'} \Big)
\end{align}
with the Fourier transform of the Coulomb potential given by $V_\q = \frac{e^2}{\epsilon\epsilon_0}\frac{1}{q^2}$, where $e$ is the elementary charge and $\epsilon_0$ 
is the vacuum permittivity.
The dielectric constant $\epsilon \approx 10$ includes the contribution of the crystal lattice\cite{Rashba_Excitons, Strzalkowski_Dielectric-constant}.
Thus, $H_\t{C}$ comprises all direct electron-electron, hole-hole, and electron-hole Coulomb interactions.

We account for the effects of a homogeneous external magnetic field $\mbf B$ on the electrons, holes, and magnetic impurity atoms, respectively, 
via the Zeeman terms
\begin{subequations}
\label{eq H_Z}
\begin{align}
H_\t{Z}^\t{e} &= \gel \mu_B \sum_{l l' \k} \mbf B \cdot \mbf s^\t{e}_{l l'} c^\dagger_{l \k} c_{l' \k},
	\an
H_\t{Z}^\t{h} &= -2 \kappa \mu_B \sum_{v v' \k} \mbf B \cdot \mbf J_{v v'} d^\dagger_{v \k} d_{v' \k},
	\an
H_\t{Z}^\t{Mn} &= \gMn \mu_B \sum_{I n n'} \mbf B \cdot \mbf S_{n n'} \hat{P}^{I}_{n n'}.
\end{align}
\end{subequations}
In the above formulas, $\gel$ is the $g$ factor of the electrons, $\kappa$ is the isotropic valence-band $g$ factor\cite{Winkler_Spin-Orbit}, 
$\gMn$ denotes the impurity $g$ factor, and $\mu_B$ is the Bohr magneton.
The vector of electron-spin matrices is given by $\mbf s^\t{e}_{l l'}$, $\mbf J_{v v'}$ is the vector of $4\times 4$ angular momentum matrices
when accounting for heavy hole (hh) and light hole (lh) bands with angular momentum $v,v' \in \{-\frac{3}{2}, -\frac{1}{2}, \frac{1}{2}, \frac{3}{2}\}$ and $\mbf S_{n n'}$ 
denotes the vector of impurity spin matrices.
In the case of manganese considered here, we have $n,n' \in \{ -\frac{5}{2}, -\frac{3}{2}, ..., \frac{5}{2} \}$.
The impurity spin itself is described by the operator $\hat{P}^{I}_{n n'} = | I,n \rangle \langle I,n' |$ where the ket $| I,n \rangle$
denotes the spin state $n$ of an impurity atom $I$.

Rather than assuming some initial carrier distribution, we explicitly account for the optical excitation and thus the light-matter coupling
via the Hamiltonian
\begin{align}
\label{eq H_lm}
H_\t{lm} = - \sum_{l v \k} \left( \mbf E \cdot \mbf M_{l v} c^\dagger_{l \k} d^\dagger_{v -\k} + \mbf E \cdot \mbf M_{v l} d_{v -\k} c_{l \k}\right)
\end{align}
with an electric field $\mbf E$ and the dipole moment $\mbf M_{l v}$ for a transition from a state in the valence subband $v$ to the conduction subband $l$.
Here, the well-known dipole approximation\cite{Rossi_Theory-of} is used to consider only interband transitions with vanishing center of mass momentum.

The dominant spin depolarization mechanism in DMS is given by the $sp$-$d$ exchange interaction which models the scattering
of $s$-like conduction-band electrons and $p$-like valence-band holes, respectively, at the localized $d$-shell electrons of the Mn impurities.
These interactions can be written as\cite{Dietl_Dilute-ferromagnetic, Kossut_Introduction-to, Cygorek_Carrier-impurity}
\begin{subequations}
\label{eq H_sd and H_pd}
\begin{align}
\label{eq H_sd}
H_{sd} &= \frac{\Jsd}{V} \sum_{\substack{I n n' \\ l l' \k \k'}} \mbf S_{n n'} \cdot \mbf s^\t{e}_{l l'} c^\dagger_{l \k} c_{l' \k'} e^{i(\k' - \k)\cdot\R_I} \! \hat{P}^{I}_{n n'},
	\an
\label{eq H_pd}
H_{pd} &= \frac{\Jpd}{V} \sum_{\substack{I n n' \\ v v' \k \k'}} \mbf S_{n n'} \cdot \mbf s^\t{h}_{v v'} d^\dagger_{v \k} d_{v' \k'} e^{i(\k' - \k)\cdot\R_I} \! \hat{P}^{I}_{n n'}
\end{align}
\end{subequations}
with the hole spin matrices given by $\mbf s^\t{h}_{v v'} = \frac{1}{3} \mbf J_{v v'}$.
Note that we employ the convention that the factor $\hbar$ which typically enters in the definition of the spin matrices is instead
absorbed in the coupling constants $\Jsd$ and $\Jpd$ as well as $\mu_B$ in case of the Zeeman terms.

In a recently published paper it was shown that the combined action of nonmagnetic impurity scattering and magnetic exchange interaction may have a significant impact
on the spin dynamics of conduction band electrons\cite{Cygorek_Influence-of}.
Therefore, we also include the nonmagnetic impurity scattering in the form
\begin{subequations}
\label{eq H_nm}
\begin{align}
\label{eq H_nm_e}
H_\t{nm}^\t{e} &= \frac{J_0^\t{e}}{V} \sum_{\substack{I l \\ \k \k'}} c^\dagger_{l \k} c_{l \k'} e^{i(\k' - \k)\cdot\R_I},
	\an
\label{eq H_nm_h}
H_\t{nm}^\t{h} &= \frac{J_0^\t{h}}{V} \sum_{\substack{I v \\ \k \k'}} d^\dagger_{v \k} d_{v \k'} e^{i(\k' - \k)\cdot\R_I}
\end{align}
\end{subequations}
with scattering constants $J_0^\t{e}$ and $J_0^\t{h}$ for electrons and holes, respectively.
Considering a DMS of the general form A$_{1-x}$Mn$_x$B, these can be determined under the assumption that unit cells containing doping ions experience an energetic 
penalty due to being forced into the same structure as the surrounding semiconductor lattice AB. 
This allows for an estimation of the nonmagnetic coupling strength based on the change of the band gap of the pure AB material compared to
the pure MnB material.
Note that we only take into account the short-range part of the carrier-impurity interaction even though it stems largely from the Coulomb interaction between the impurity atoms
and the quasi-free carriers\cite{Cygorek_Influence-of}.

We do not include the influence of phonons on the carrier spin dynamics in our model since typical experiments\cite{BenCheikh_Electron-spin, Camilleri_Electron-and, Vladimirova_Dynamics-of} 
are performed at low temperatures of about $2\,$K where only phonon emission is relevant because there are no phonons available for absorption.
But since we consider only direct laser-driven excitation of excitons with vanishing center of mass momenta, phonon emission processes are also strongly suppressed as there are no 
final exciton states lower in energy to scatter to.
Additionally, phonons do not couple directly to the spin and thus represent a secondary relaxation process which only becomes relevant in combination with other effects, such as
spin-orbit coupling.
Theoretical rate-equation models that include the scattering due to phonons also support that the $s$-$d$ exchange interaction is the most important scattering mechanism at low 
temperatures\cite{Jiang_Electron-spin}.
Given that the recently reported\cite{Debus_Resonantly-enhanced} spin-lattice relaxation time of Mn$^{2+}$ ions in typical DMS quantum wells is on the order of $\mu$s,
the coupling of phonons to the Mn system can also be disregarded on the typical ps time scale of the carrier spin relaxation\cite{Cygorek_Influence-of, Ungar_Ultrafast-spin, 
BenCheikh_Electron-spin}.
Furthermore, spin-orbit effects\cite{Ungar_Ultrafast-spin} as well as the hyperfine interaction\cite{Wu_Spin-dynamics} due to nuclear spins typically also only become relevant
at much longer time scales.

In a quantum well, it is convenient to switch from a three-dimensional basis set to a description where only the in-plane part consists of plane waves and the $z$ dependence 
is treated separately.
One can then expand the single-particle basis functions $\Xi$ in terms of a complete set of envelope functions, which yields
\begin{align}
\Xi(\r, z) = \frac{1}{\sqrt{A}}\sum_{n \k} a_{n \k} e^{i\k\cdot\r} \, u_n^\t{e/h}(z)
\end{align}
with envelope functions $u_n^\t{e/h}(z)$ of electrons and holes, respectively, and expansion coefficients $a_{n \k}$.
Here and throughout the remainder of this article, the appearing wave vectors $\k$ as well as the in-plane position $\r$ are two-dimensional quantities.

For narrow quantum wells, where the energetic separation between the individual confinement states is large, it is a good approximation to only consider the lowest confinement
state\cite{Astakhov_Binding-energy} $u_0^\t{e/h}(z)$, which corresponds to setting $a_{n \k} = 0$ for all  $n \neq 0$.
Thus, we project the Hamiltonian given by Eq.~\eqref{eq complete Hamiltonian} onto the corresponding subspace.
For the carrier-impurity interactions in Eqs.~\eqref{eq H_sd and H_pd} and \eqref{eq H_nm}, this amounts to substituting $\sum_{k_z k_z'} \to d |u_0^\t{e/h}(Z_I)|^2$.
In numerical calculations, we assume infinitely high potential barriers at $z = \pm \frac{d}{2}$, so that the envelope functions for electrons and holes become
\begin{align}
\label{eq envelope function approximation}
u_0^\t{e/h}(z) = \sqrt{\frac{2}{d}}\cos\Big(\frac{\pi}{d}z\Big).
\end{align}

\subsection{Dynamical variables and truncation scheme}
\label{subsec Dynamical variables and truncation scheme}

Our main target is the modeling of the electron or hole spin dynamics in a system where all particles are excited optically as electron-hole pairs.
Within the DCT scheme this is most conveniently achieved by deriving quantum kinetic equations of motion for the four-point density matrices
$\langle c^\dagger_{l_1 \k_1} d^\dagger_{v_1 \k_2} d_{v_2 \k_3} c_{l_2 \k_4} \rangle$ from which all relevant information can be deduced\cite{Axt_Nonlinear-optics}.
To provide an example, the electron density matrix is given by
\begin{align}
\langle c^\dagger_{l_1 \k_1} c_{l_2 \k_2} \rangle &= \sum_{v \k} \langle c^\dagger_{l_1 \k_1} d^\dagger_{v \k} d_{v \k} c_{l_2 \k_2} \rangle + \mathcal{O}(\mbf E^4).
\end{align}
Starting from the Hamiltonian given by Eq.~\eqref{eq complete Hamiltonian} and using the Heisenberg equation of motion, one ends up with an infinite hierarchy of
equations that needs to be truncated in order to be solvable.
In this article, we employ a dynamics-controlled truncation\cite{Axt_A-dynamics} which classifies all appearing expectation values in terms of their order in the
generating optical field.
Using this procedure, we keep all contributions up to the order $\mathcal{O}(\mbf E^2)$, which is sufficient in the low-density regime\cite{Siantidis_Dynamics-of}.

However, since we are dealing with a DMS, we also have to treat correlations between carriers and Mn atoms.
This is done using a correlation expansion similarly to Ref.~\onlinecite{Thurn_Quantum-kinetic} where, due to the Mn atoms being far apart in a DMS, correlations
that involve magnetic dopants at different sites are disregarded.
Applications of correlation expansions in condensed matter physics are manifold and can be found explained numerous times in the literature\cite{Zimmermann_Non-Markovian,
Rossi_Theory-of, Thurn_Quantum-kinetic, Kira_Cluster-expansion, Kira_Many-body, Kuhn_Density-matrix}.

Setting up the equations of motion for an on-average spatially homogeneous system, a closed set of equations of motion can be formulated for the following dynamical variables:
\begin{widetext}
\begin{subequations}
\label{eq dynamical variables}
\begin{align}
M_{n_1}^{n_2}(z) &= \frac{d}{\NMn} \sum_I \delta(z-Z_I) \big\langle \hat{P}_{n_1 n_2}^{I} \big\rangle,
	\an
Y_{\k_1}^{v_1 l_1} &= \big\langle d_{v_1 -\k_1} c_{l_1 \k_1} \big\rangle,
	\an
\label{eq dynamical variables N}
N_{\k_1 \k_2 \k_3 \k_4}^{l_1 v_1 v_2 l_2} &= \big\langle c^\dagger_{l_1 \k_1} d^\dagger_{v_1 -\k_2} d_{v_2 -\k_3} c_{l_2 \k_4} \big\rangle 
	\delta_{\k_1-\k_2,\k_4-\k_3},
	\an
{Q_Y}_{n_1 \k_1 \k_2}^{n_2 v_1 l_1}(z) &= V \frac{d}{\NMn} \sum_I \delta(z-Z_I) \big\langle d_{v_1 -\k_1} c_{l_1 \k_2} e^{i(\k_2-\k_1)\cdot\R_{I}} \hat{P}_{n_1 n_2}^{I} \big\rangle 
	\t{, for } \k_1 \neq \k_2,
	\an
\bar{Y}_{\k_1 \k_2}^{v_1 l_1}(z) &= V \frac{d}{\NMn} \sum_I \delta(z-Z_I) \big\langle d_{v_1 -\k_1} c_{l_1 \k_2} e^{i(\k_2-\k_1)\cdot\R_{I}} \big\rangle \t{, for } \k_1 \neq \k_2,
	\an
{Q_N}_{n_1 \k_1 \k_2 \k_3 \k_4}^{n_2 l_1 v_1 v_2 l_2}(z) &= V \frac{d}{\NMn} \sum_I \delta(z-Z_I) \big\langle c^\dagger_{l_1 \k_1} d^\dagger_{v_1 -\k_2} d_{v_2 -\k_3} c_{l_2 \k_4} 
	e^{i(\k_2-\k_1+\k_4-\k_3)\cdot\R_{I}} \hat{P}_{n_1 n_2}^{I} \big\rangle \t{, for } \k_1-\k_2 \neq \k_4-\k_3,
	\an
\bar{N}_{\k_1 \k_2 \k_3 \k_4}^{l_1 v_1 v_2 l_2} &= V \frac{d}{\NMn} \sum_I \delta(z-Z_I) \big\langle c^\dagger_{l_1 \k_1} d^\dagger_{v_1 -\k_2} d_{v_2 -\k_3} c_{l_2 \k_4}
	e^{i(\k_2-\k_1+\k_4-\k_3)\cdot\R_{I}} \big\rangle \t{, for } \k_1-\k_2 \neq \k_4-\k_3.
\end{align}
\end{subequations}
\end{widetext}

In the above equations, $M_{n_1}^{n_2}(z)$, $Y_{\k_1}^{v_1 l_1}$, and $N_{\k_1 \k_2 \k_3 \k_4}^{l_1 v_1 v_2 l_2}$ represent the Mn density matrices, the electron-hole coherences, 
and the exciton density matrices, respectively.
The magnetic and nonmagnetic correlations between coherences and impurity atoms are given by ${Q_Y}_{n_1 \k_1 \k_2}^{n_2 v_1 l_1}(z)$ and $\bar{Y}_{\k_1 \k_2}^{v_1 l_1}(z)$,
respectively, and in turn by ${Q_N}_{n_1 \k_1 \k_2 \k_3 \k_4}^{n_2 l_1 v_1 v_2 l_2}(z)$ and $\bar{N}_{\k_1 \k_2 \k_3 \k_4}^{l_1 v_1 v_2 l_2}$ between excitons and impurities.
In addition to the usual quantum mechanical average of the operators, the brackets $\langle . \rangle$ in Eqs.~\eqref{eq dynamical variables} as well as throughout the rest of this paper
also contain an average over the distribution of Mn positions in the sample.
This distribution is assumed to be random but homogeneous on average, so that $\langle e^{i(\k_2-\k_1)\cdot\R_{I}} \rangle = \delta_{\k_1,\k_2}$.
The delta distribution in Eq.~\eqref{eq dynamical variables N} is a consequence of the spatial homogeneity of the system.

Using these variables, it is straightforward but lengthy to set up a hierarchy of equations of motion whilst retaining only terms up to $\mathcal{O}(\mbf E^2)$ according to the DCT scheme.
However, it turns out that the magnetic interactions $H_{sd}$ and $H_{pd}$ introduce additional source terms in the equations for the correlations that are not expressible using the 
variables from Eqs.~\eqref{eq dynamical variables} because they contain products of Mn operators as well as exponential functions containing the randomly distributed Mn positions
$\R_I$ in the exponent.
Following along the lines of Ref.~\onlinecite{Thurn_Quantum-kinetic}, where a correlation expansion has been successfully employed to treat these terms, we sketch the general method
of such an expansion when applied to the expressions derived in this paper.
Our approach for dealing with random impurity positions can also be related to the treatment of interface roughness via random potentials as well as the influence of disorder 
in semiconductors\cite{Zimmermann_Exciton-lineshape, Zimmermann_Theory-of-1992, Zimmermann_Theory-of-1995}.

Consider a general expectation value of the form 
\begin{align}
\label{eq general form of source term for Q}
S_Q &= \langle X e^{i\Delta\k\cdot\R_{I}} e^{i\Delta\k'\cdot\R_{I'}} \hat{P}_{n_1 n_2}^{I} \hat{P}_{n_1' n_2'}^{I'} \rangle,
\end{align}
where $X$ contains up to four Fermi operators so that $\langle X \rangle$ is up to $\mathcal{O}(E^2)$ and $\Delta\k \neq \0$.
Using the DCT scheme, it can be easily shown that assisted expectation values such as the quantity in Eq.~\eqref{eq general form of source term for Q} are of the same order in the generating
electric field as the corresponding bare expectation values of the Fermi operators.
We then treat the expression in Eq.~\eqref{eq general form of source term for Q} as follows:

(i) The situation $I = I'$ has to be considered separately since in this case we are dealing with Mn operators on the same site, so that Eq.~\eqref{eq general form of source term for Q}
reduces to
\begin{align}
S_Q \stackrel{I = I'}{=} \langle X e^{i(\Delta\k+\Delta\k')\cdot\R_{I}} \hat{P}_{n_1 n_2'}^{I} \rangle \delta_{n_2,n_1'}
\end{align}
in accordance with the definition of the Mn operators $\hat{P}_{n n'}^I$.
The remaining quantity can then be expressed in terms of the variables introduced in Eq.~\eqref{eq dynamical variables}.

(ii) If $\Delta\k' = \0$, we get 
\begin{align}
S_Q \stackrel{\Delta\k' = \0}{=} \langle X e^{i\Delta\k\cdot\R_{I}} \hat{P}_{n_1 n_2}^{I} \hat{P}_{n_1' n_2'}^{I'} \rangle,
\end{align}
so that the number of operators effectively is reduced by one.

(iii) In the most general case, i.e., $I \neq I'$ and $\Delta\k' \neq 0$, we decompose Eq.~\eqref{eq general form of source term for Q} using a correlation expansion.
This yields
\begin{align}
S_Q &= \delta\langle X e^{i\Delta\k\cdot\R_{I}} \hat{P}_{n_1 n_2}^{I} \rangle \langle e^{i\Delta\k'\cdot\R_{I'}} \rangle \langle \hat{P}_{n_1' n_2'}^{I'} \rangle
	\nn
	&\phantom{=\;} + \delta\langle X e^{i\Delta\k\cdot\R_{I}} \rangle \langle \hat{P}_{n_1 n_2}^{I} \rangle \langle e^{i\Delta\k'\cdot\R_{I'}} \rangle \langle \hat{P}_{n_1' n_2'}^{I'} \rangle
\end{align}
with true correlations denoted by $\delta\langle ... \rangle$.
In the above equation, we have only written down the non-vanishing terms of the expansion by neglecting correlations evaluated either at different Mn sites or involving
two or more impurity operators.
Furthermore, it can be shown that correlations of the form $\delta\langle e^{i\Delta\k\cdot\R_{I}} \hat{P}_{n_1 n_2}^{I} \rangle$, which could be used to model impurity spin waves,
are not driven during the dynamics if they are zero initially and thus need not be explicitly accounted for.

This approach enables the formulation of a closed set of equations of motion containing only reduced density matrices and the true correlations.
However, instead of using the true correlations as dynamical variables, we switch back to the non-factorized correlations [c.f. Eqs.~\eqref{eq dynamical variables}] because this
allows for a much more condensed and convenient notation of the equations of motion.

\subsection{Transformation to the exciton basis}
\label{subsec Transformation to the exciton basis}

Since the highest-order density matrices depend on four wave vectors, the resulting equations are numerically very demanding.
Instead, when essentially only bound excitons are excited, it is much more convenient and efficient to use a two-particle 
basis\cite{Dresselhaus_Effective-mass, Axt_Nonlinear-optics, Siantidis_Dynamics-of, Iotti_A-model, Winkler_Excitons-and}, which in this case allows for a significant reduction
of relevant basis states.
We note in passing that one could also change to the exciton basis before deriving equations of motion.
However, this way a classification of contributions to the equations of motion in terms of powers of the electric field is not straightforward.
Therefore, we first derive the equations of motion in the single-particle basis and transform to the two-particle basis afterwards.

We consider the excitonic eigenvalue problem in the quantum well plane given by 
\begin{align}
\label{eq Exciton problem}
\big( H_0^{\t{e}} + H_0^{\t{h}} + H_\t{C} \big) \psi_{x \K}(\r^\t{e}, \r^\t{h}) = E_{x \K} \psi_{x \K}(\r^\t{e}, \r^\t{h})
\end{align}
with the exciton energy $E_{x \K}$ and the two-dimensional position vectors of the electron and the hole $\r^\t{e}$ and $\r^\t{h}$, respectively.  
Using the effective mass approximation as well as the strong confinement limit of the Coulomb interaction, the Hamiltonians read
\begin{subequations}
\begin{align}
H_0^\t{e} &= -\frac{\hbar^2}{2 \me}(\partial_{x^\t{e}}^2 + \partial_{y^\t{e}}^2) + E_\t{g},
	\an
H_0^\t{h} &= -\frac{\hbar^2}{2 \mh}(\partial_{x^\t{h}}^2 + \partial_{y^\t{h}}^2),
	\an
\label{eq effective Coulomb Hamiltonian}
H_\t{C} &= - \int \! dz \int \! dz' \frac{e^2 |u_0^\t{e}(z)|^2 |u_0^\t{h}(z')|^2}{4\pi\epsilon\epsilon_0\sqrt{(\r^\t{e} \! - \! \r^\t{h})^2 \! + \! (z \! - \! z')^2}}
\end{align}
\end{subequations}
with in-plane electron and heavy hole effective masses $\me$ and $\mh$, respectively, as well as the band gap $E_\t{g}$.
The exciton wave function can be decomposed into a center of mass and a relative part according to
\begin{align}
\psi_{x \K}(\r^\t{e}, \r^\t{h}) = \frac{1}{\sqrt{A}} e^{i\K\cdot\R} \phi_x(\r)
\end{align}
with the exciton center of mass momentum $\K$ and the exciton quantum number $x$.
The relative coordinate is given by $\r = \r^\t{e} - \r^\t{h}$ and $\R = \etae\r^\t{e} + \etah\r^\t{h}$ denotes the center of mass coordinate of the exciton with the mass ratios 
$\etae := \frac{\me}{M}$ and $\etah := \frac{\mh}{M}$, where $M = \me+\mh$ is the exciton mass.

Using polar coordinates, the relative part of the exciton wave function in two dimensions can be further decomposed into a radial part $R_n(r)$ with a principal quantum 
number $n \in \mathbb{N}$ and an angular part $e^{il\varphi}$ with angular momentum quantum number $|l| = 0, 1, ..., n-1$ according to\cite{Bastard_Wave-mechanics, Yang_Analytic-solution}
\begin{align}
\label{eq decomposition of wave function}
\phi_{x}(\r) = R_{n}(r) e^{il\varphi},
\end{align}
where the quantum numbers $n$ and $l$ are condensed into a single index $x$.

The creation operator of an exciton with an electron in the conduction band $l$ and a hole in the valence band $v$ can be written as
\begin{align}
{{\hat Y}^\dagger}_{l v x \K} = \int \! d^2 r^\t{e} \int \! d^2 r^\t{h} \, \psi_{x \K}(\r^\t{e}, \r^\t{h}) \, \hat{\chi}_{l \r^\t{e}}^\dagger \, \hat{\chi}_{v \r^\t{h}}
\end{align}
using the Wannier operators
\begin{subequations}
\begin{align}
\hat{\chi}_{l \r^\t{e}}^\dagger &= \frac{1}{\sqrt{A}} \sum_{\k^\t{e}} e^{-i\k^\t{e}\cdot\r^\t{e}} c^\dagger_{l \k^\t{e}},
	\an
\hat{\chi}_{v \r^\t{h}} &= \frac{1}{\sqrt{A}} \sum_{\k^\t{h}} e^{-i\k^\t{h}\cdot\r^\t{h}} d^\dagger_{v \k^\t{h}}.
\end{align}
\end{subequations}
Then, the relation between the exciton creation operator and the Fermi operators reads
\begin{subequations}
\begin{align}
\label{eq trafo between excitons and Fermi operators}
{{\hat Y}^\dagger}_{l v x \K} &= \sum_{\k^\t{e} \k^\t{h}} \psi_{x \K}^{\k^\t{e}, -\k^\t{h}} \, c^\dagger_{l \k^\t{e}} d^\dagger_{v -\k^\t{h}},
	\an
c^\dagger_{l \k^\t{e}} d^\dagger_{v -\k^\t{h}} &= \sum_{x \K} \big(\psi_{x \K}^{\k^\t{e},-\k^\t{h}}\big)^* \, {{\hat Y}^\dagger}_{l v x \K}
\end{align}
\end{subequations}
with the matrix element
\begin{align}
\psi_{x \K}^{\k^\t{e} \k^\t{h}} &:= \frac{1}{\sqrt{A}} \delta_{\K, \k^\t{e}+\k^\t{h}} \int \! d^2r \, e^{-i\r\cdot(\etah\k^\t{e} - \etae\k^\t{h})} \phi_x(\r).
\end{align}

Using the transformation in Eq.~\eqref{eq trafo between excitons and Fermi operators}, we switch from the quantities defined in Eqs.~\eqref{eq dynamical variables}
to the new dynamical variables in the exciton basis
\begin{subequations}
\label{eq trafo to exciton basis}
\begin{align}
Y_{x_1 \0}^{v_1 l_1} &= \sum_{\k_1} \big(\psi_{x_1 \0}^{\k_1,-\k_1}\big)^* \, Y_{\k_1}^{v_1 l_1},
	\an
N_{x_1 \K_1}^{l_1 v_1 v_2 l_2} &= \sum_{\substack{\k_1 \k_2 \\ \k_3 \k_4}} \psi_{x_1 \K_1}^{\k_1,-\k_2} \big(\psi_{x_1 \K_1}^{\k_4,-\k_3}\big)^* \, N_{\k_1 \k_2 \k_3 \k_4}^{l_1 v_1 v_2 l_2},
	\an
{Q_Y}_{n_1 x_1 \K_1}^{n_2 v_1 l_1}(z) &= \sum_{\k_1 \k_2} \big(\psi_{x_1 \K_1}^{\k_2,-\k_1}\big)^* \, {Q_Y}_{n_1 \k_1 \k_2}^{n_2 v_1 l_1}(z),
	\an
\bar{Y}_{x_1 \K_1}^{v_1 l_1}(z) &= \sum_{\k_1 \k_2} \big(\psi_{x_1 \K_1}^{\k_2,-\k_1}\big)^* \, \bar{Y}_{\k_1 \k_2}^{v_1 l_1}(z),
	\an
{Q_N}_{n_1 x_1 \K_1 x_2 \K_2}^{n_2 l_1 v_1 v_2 l_2}(z) &= \sum_{\substack{\k_1 \k_2 \\ \k_3 \k_4}} \psi_{x_1 \K_1}^{\k_1,-\k_2} \big(\psi_{x_2 \K_2}^{\k_4,-\k_3}\big)^*
	\nn
	&\phantom{=\sum_{\substack{\k_1 \k_2 \\ \k_3 \k_4}}\;} \times {Q_N}_{n_1 \k_1 \k_2 \k_3 \k_4}^{n_2 l_1 v_1 v_2 l_2}(z),
	\an
\bar{N}_{x_1 \K_1 x_2 \K_2}^{l_1 v_1 v_2 l_2}(z) &= \! \sum_{\substack{\k_1 \k_2 \\ \k_3 \k_4}} \! \psi_{x_1 \K_1}^{\k_1,-\k_2} \big(\psi_{x_2 \K_2}^{\k_4,-\k_3}\big)^* 
	\bar{N}_{\k_1 \k_2 \k_3 \k_4}^{l_1 v_1 v_2 l_2}(z).
\end{align}
\end{subequations}

\subsection{Equations of motion}
\label{subsec Equations of motion}

Applying the DCT scheme and the correlation expansion in the equations of motion in the electron-hole representation and subsequently using the transformation to the exciton basis
according to Eqs.~\eqref{eq trafo to exciton basis} leads to the following equations of motion:
\begin{widetext}
\begin{subequations}
\label{eq full equations of motion}
\begin{align}
\label{eq EOM for Mn}
&i\hbar\ddt M_{n_1}^{n_2}(z)
	= \sum_{n} \Big( \mbf S_{n_2 n} M_{n_1}^{n}(z) - \mbf S_{n n_1} M_{n}^{n_2}(z) \Big) \cdot 
	\Big( \hbar\bs\omMn + \frac{\Jsd}{V} d |u_0^\t{e}(z)|^2 \sum_{\substack{l l' v\\ x \K}} \mbf s^\t{e}_{l l'} N_{x \K}^{l v v l'}
	+ \frac{\Jpd}{V} d |u_0^\t{h}(z)|^2 \sum_{\substack{v v' l \\ x \K}} \mbf s^\t{h}_{v v'} N_{x \K}^{l v v' l} \Big)
	\nn
	&\phantom{i\hbar\ddt M_{n_1}^{n_2}(z)=\;} + \frac{\Jsd}{V^2} d |u_0^\t{e}(z)|^2 \!\! \sum_{\substack{l l' v n \\ x \K x' \K'}} \!\! \mbf s^\t{e}_{l l'} \cdot 
	f_{-\etah x x'}^{\phantom{-\etah} \K \K'} \Big( \mbf S_{n_2 n} {Q_N}_{n_1 x \K x' \K'}^{n l v v l'}(z) - \mbf S_{n n_1} {Q_N}_{n x \K x' \K'}^{n_2 l v v l'}(z) \Big)
	\nn
	&\phantom{i\hbar\ddt M_{n_1}^{n_2}(z)=\;} + \frac{\Jpd}{V^2} d |u_0^\t{h}(z)|^2 \!\! \sum_{\substack{v v' l n \\ x \K x' \K'}} \!\! \mbf s^\t{h}_{v v'} \cdot
	f_{\etae x x'}^{\phantom{\etae} \K \K'} \Big( \mbf S_{n_2 n} {Q_N}_{n_1 x \K x' \K'}^{n l v v' l}(z) - \mbf S_{n n_1} {Q_N}_{n x \K x' \K'}^{n_2 l v v' l}(z) \Big),
	\an
\label{eq EOM for Y}
&i\hbar\ddt Y_{x_1 \0}^{v_1 l_1}
	= - \mbf E \cdot \mbf M_{l_1 v_1}^{x_1}
	+ \Big(E_{x_1 \0}^{v_1 l_1} + \frac{(J_0^\t{e} \!+\! J_0^\t{h}) \NMn}{V} \Big) Y_{x_1 \0}^{v_1 l_1}
	+ \sum_{l} \hbar\bs\ome \cdot \mbf s^\t{e}_{l_1 l} Y_{x_1 \0}^{v_1 l}
	+ \sum_{v} \hbar\bs\omh \cdot \mbf s^\t{h}_{v_1 v} Y_{x_1 \0}^{v l_1}
	\nn
	&\phantom{i\hbar\ddt Y_{x_1 \0}^{v_1 l_1}=\;} 
	+ \frac{\NMn}{V^2} \sum_{\substack{n n' \\ x \K}} \mbf S_{n n'} \cdot\! \int \! dz 
	\Big( \Jsd |u_0^\t{e}(z)|^2 \sum_{l} \mbf s^\t{e}_{l_1 l} f_{-\etah x_1 x}^{\phantom{-\etah} \0 \K} {Q_Y}_{n x \K}^{n' v_1 l}(z)
	+ \Jpd |u_0^\t{h}(z)|^2 \sum_{v} \mbf s^\t{h}_{v_1 v} f_{\etae x_1 x}^{\phantom{\etae} \0 \K} {Q_Y}_{n x \K}^{n' v l_1}(z) \Big)
	\nn
	&\phantom{i\hbar\ddt Y_{\0 x_1}^{v_1 l_1}=\;} 
	+ \frac{\NMn}{V^2} \sum_{x \K} \int \! dz \Big( J_0^\t{e} |u_0^\t{e}(z)|^2 f_{-\etah x_1 x}^{\phantom{-\etah} \0 \K} \bar{Y}_{x \K}^{v_1 l_1}(z)
	+ J_0^\t{h} |u_0^\t{h}(z)|^2 f_{\etae x_1 x}^{\phantom{\etae} \0 \K} \bar{Y}_{x \K}^{v_1 l_1}(z) \Big),
	\an
\label{eq EOM for N}
&i\hbar\ddt N_{x_1 \K_1}^{l_1 v_1 v_2 l_2}
	= \mbf E \cdot \Big( \mbf M_{v_1 l_1}^{x_1} Y_{x_1 \0}^{v_2 l_2} - \mbf M_{l_2 v_2}^{x_1} \big(Y_{x_1 \0}^{v_1 l_1}\big)^* \Big) \delta_{\K_1,\0}
	+ \Big(E_{x_1 \K_1}^{v_2 l_2} - E_{x_1 \K_1}^{v_1 l_1}\Big) N_{x_1 \K_1}^{l_1 v_1 v_2 l_2}
	\nn
	&\phantom{i\hbar\ddt N_{x_1 \K_1}^{l_1 v_1 v_2 l_2}=\;} 
	+ \sum_{l} \hbar\bs\ome \cdot \Big( \mbf s^\t{e}_{l_2 l} N_{x_1 \K_1}^{l_1 v_1 v_2 l} 
	- \mbf s^\t{e}_{l l_1} N_{x_1 \K_1}^{l v_1 v_2 l_2} \Big)
	+ \sum_{v} \hbar\bs\omh \cdot \Big( \mbf s^\t{h}_{v_2 v} N_{x_1 \K_1}^{l_1 v_1 v l_2} 
	- \mbf s^\t{h}_{v v_1} N_{x_1 \K_1}^{l_1 v v_2 l_2} \Big)
	\nn
	&\phantom{i\hbar\ddt N_{x_1 \K_1}^{l_1 v_1 v_2 l_2}=\;} 
	+ \frac{\Jsd \NMn}{V^2} \! \int \! dz |u_0^\t{e}(z)|^2 \sum_{\substack{l n n' \\ x \K}} \mbf S_{n n'} \cdot \Big(
	\mbf s^\t{e}_{l_2 l} f_{-\etah x_1 x}^{\phantom{-\etah} \K_1 \K} {Q_N}_{n x_1 \K_1 x \K}^{n' l_1 v_1 v_2 l}(z)
	- \mbf s^\t{e}_{l l_1} f_{-\etah x x_1}^{\phantom{-\etah} \K \K_1} {Q_N}_{n x \K x_1 \K_1}^{n' l v_1 v_2 l_2}(z) \Big)
	\nn
	&\phantom{i\hbar\ddt N_{x_1 \K_1}^{l_1 v_1 v_2 l_2}=\;} 
	+ \frac{\Jpd \NMn}{V^2} \! \int \! dz |u_0^\t{h}(z)|^2 \sum_{\substack{v n n' \\ x \K}} \mbf S_{n n'} \cdot \Big(
	\mbf s^\t{h}_{v_2 v} f_{\etae x_1 x}^{\phantom{\etae} \K_1 \K} {Q_N}_{n x_1 \K_1 x \K}^{n' l_1 v_1 v l_2}(z)
	- \mbf s^\t{h}_{v v_1} f_{\etae x x_1}^{\phantom{\etae} \K \K_1} {Q_N}_{n x \K x_1 \K_1}^{n' l_1 v v_2 l_2}(z) \Big)
	\nn
	&\phantom{i\hbar\ddt N_{x_1 \K_1}^{l_1 v_1 v_2 l_2}=\;} 
	+ \frac{J_0^\t{e} \NMn}{V^2} \! \int \! dz |u_0^\t{e}(z)|^2 \sum_{x \K} 
	\Big( f_{-\etah x_1 x}^{\phantom{-\etah} \K_1 \K} \bar{N}_{x_1 \K_1 x \K}^{l_1 v_1 v_2 l_2}(z) 
	- f_{-\etah x x_1}^{\phantom{-\etah} \K \K_1} \bar{N}_{x \K x_1 \K_1}^{l_1 v_1 v_2 l_2}(z) \Big)
	\nn
	&\phantom{i\hbar\ddt N_{x_1 \K_1}^{l_1 v_1 v_2 l_2}=\;} 
	+ \frac{J_0^\t{h} \NMn}{V^2} \! \int \! dz |u_0^\t{h}(z)|^2 \sum_{x \K} 
	\Big( f_{\etae x_1 x}^{\phantom{\etae} \K_1 \K} \bar{N}_{x_1 \K_1 x \K}^{l_1 v_1 v_2 l_2}(z) 
	- f_{\etae x x_1}^{\phantom{\etae} \K \K_1} \bar{N}_{x \K x_1 \K_1}^{l_1 v_1 v_2 l_2}(z) \Big),
	\an
\label{eq EOM for Q_Y}
&i\hbar\ddt {Q_Y}_{n_1 x_1 \K_1}^{n_2 v_1 l_1}(z)
	= \Big(E_{x_1 \K_1}^{v_1 l_1} + \frac{(J_0^\t{e} \! + \! J_0^\t{h}) \NMn}{V}\Big) {Q_Y}_{n_1 x_1 \K_1}^{n_2 v_1 l_1}(z)
	+ {\beta_{n_1 x_1 \K_1}^{n_2 v_1 l_1}(z)}^{I}
	+ {\beta_{n_1 x_1 \K_1}^{n_2 v_1 l_1}(z)}^{II}
	+ {\beta_{n_1 x_1 \K_1}^{n_2 v_1 l_1}(z)}^{III},
	\an
\label{eq EOM for Y_bar}
&i\hbar\ddt \bar{Y}_{x_1 \K_1}^{v_1 l_1}(z)
	= \Big(E_{x_1 \K_1}^{v_1 l_1} + \frac{(J_0^\t{e} \! + \! J_0^\t{h}) \NMn}{V}\Big) \bar{Y}_{x_1 \K_1}^{v_1 l_1}(z)
	+ {\bar{\beta}_{x_1 \K_1}^{v_1 l_1}(z)}^{I}
	+ {\bar{\beta}_{x_1 \K_1}^{v_1 l_1}(z)}^{II}
	+ {\bar{\beta}_{x_1 \K_1}^{v_1 l_1}(z)}^{III},
	\an
\label{eq EOM for Q_N}
&i\hbar\ddt {Q_N}_{n_1 x_1 \K_1 x_2 \K_2}^{n_2 l_1 v_1 v_2 l_2}(z)
	= \mbf E \cdot \Big( \mbf M_{v_1 l_1}^{x_1} {Q_Y}_{n_1 x_2 \K_2}^{n_2 v_2 l_2}(z) \delta_{\K_1,\0}
	- \mbf M_{l_2 v_2}^{x_2} \big({Q_Y}_{n_1 x_1 \K_1}^{n_2 v_1 l_1}(z)\big)^* \delta_{\K_2,\0} \Big)
	+ \Big(E_{x_2 \K_2}^{v_2 l_2} - E_{x_1 \K_1}^{v_1 l_1}\Big)
	\nn
	&\phantom{i\hbar\ddt {Q_N}_{n_1 x_1 \K_1 x_2 \K_2}^{n_2 l_1 v_1 v_2 l_2}(z)=\;} 
	\times {Q_N}_{n_1 x_1 \K_1 x_2 \K_2}^{n_2 l_1 v_1 v_2 l_2}(z) 
	+ {b_{n_1 x_1 \K_1 x_2 \K_2}^{n_2 l_1 v_1 v_2 l_2}(z)}^{I}
	+ {b_{n_1 x_1 \K_1 x_2 \K_2}^{n_2 l_1 v_1 v_2 l_2}(z)}^{II} 
	+ {b_{n_1 x_1 \K_1 x_2 \K_2}^{n_2 l_1 v_1 v_2 l_2}(z)}^{III},
	\an
\label{eq EOM for N_bar}
&i\hbar\ddt \bar{N}_{x_1 \K_1 x_2 \K_2}^{l_1 v_1 v_2 l_2}(z)
	= \mbf E \cdot \Big( \mbf M_{v_1 l_1}^{x_1} \bar{Y}_{x_2 \K_2}^{v_2 l_2}(z) \delta_{\K_1,0}
	- \mbf M_{l_2 v_2}^{x_2} \big(\bar{Y}_{x_1 \K_1}^{v_1 l_1}(z)\big)^* \delta_{\K_2,0} \Big)
	+ \Big(E_{x_2 \K_2}^{v_2 l_2} - E_{x_1 \K_1}^{v_1 l_1}\Big) \bar{N}_{x_1 \K_1 x_2 \K_2}^{l_1 v_1 v_2 l_2}(z)
	\nn 
	&\phantom{i\hbar\ddt \bar{N}_{x_1 \K_1 x_2 \K_2}^{l_1 v_1 v_2 l_2}(z)=\;} 
	+ {\bar{b}_{x_1 \K_1 x_2 \K_2}^{l_1 v_1 v_2 l_2}(z)}^{I}
	+ {\bar{b}_{x_1 \K_1 x_2 \K_2}^{l_1 v_1 v_2 l_2}(z)}^{II}
	+ {\bar{b}_{x_1 \K_1 x_2 \K_2}^{l_1 v_1 v_2 l_2}(z)}^{III}.
\end{align}
\end{subequations}
\end{widetext}

The mean-field precession frequencies and directions of impurities, electrons, and holes, respectively, are given by 
\begin{subequations}
\label{eq mean-field frequencies}
\begin{align}
\bs\omMn &= \frac{1}{\hbar} \gMn \mu_B \mbf B,
	\an
\bs\ome &= \frac{1}{\hbar} \gel \mu_B \mbf B + \frac{\Jsd \NMn}{\hbar V} \int dz |u_0^\t{e}(z)|^2 \langle\mbf S(z)\rangle,
	\an
\bs\omh &= - \frac{6}{\hbar} \kappa \mu_B \mbf B + \frac{\Jpd \NMn}{\hbar V} \int dz |u_0^\t{h}(z)|^2 \langle\mbf S(z)\rangle,
\end{align}
\end{subequations}
where $\langle \mbf S(z)\rangle = \sum_{n n'} \langle \mbf S_{n n'} M_{n}^{n'}(z) \rangle$ is the mean impurity spin.
In the exciton representation, the dipole matrix element becomes $\mbf M_{l v}^{x} := \mbf M_{l v} \phi_{x}(\r = \0)$.
The wave-vector dependent form factors that arise in Eqs.~\eqref{eq full equations of motion} are given by
\begin{align}
\label{eq exciton form factors}
f_{\eta x_1 x_2}^{\phantom{\eta} \K_1 \K_2} &:= \! \int \! d^2r\, e^{-i\eta (\K_1 - \K_2)\cdot\r} \phi_{x_1}^*(\r) \phi_{x_2}(\r)
	\nn
	&= 2\pi \!\! \int_0^\infty \!\!\!\! dr \, r R_{n_1}(r) R_{n_2}(r) i^{-\Delta l} e^{i\Delta l \psi_{12}} J_{\Delta l}\big( \eta K_{12} r \big)
\end{align}
with $\eta \in \{-\etah,\etae\}$, $\Delta l = l_2 - l_1$, $K_{12} = |\K_1-\K_2|$, and $J_{\Delta l}(x)$ denoting the cylindrical Bessel function of integer order $\Delta l$.
Furthermore, $\psi_{12}$ is the angle between the vector $(\K_1 - \K_2)$ and the $x$ axis.
To arrive at the above formula, the Jacobi-Anger expansion has been used.
The source terms $\beta$, $\bar{\beta}$, $b$, and $\bar{b}$ for the correlations are listed in Eqs.~\eqref{eq source terms for the correlations} in the appendix.

In the equations of motion, one can identify terms with different physical interpretation.
For instance, in Eq.~\eqref{eq EOM for Y}, the first term on the right-hand side represents the optical driving by the laser field, followed by a homogeneous term proportional 
to the quasiparticle energy of the exciton.
Note that the nonmagnetic impurity interaction renormalizes the band gap and therefore the quasiparticle energy.
The terms proportional to $\bs\ome$ and $\bs\omh$ describe the precession around the effective field due to the external magnetic field as well as the impurity magnetization.
The influence of the magnetic carrier-impurity correlations is given by the terms proportional to the magnetic coupling constants $\Jsd$ and $\Jpd$, while terms proportional to 
$J_0^\t{e}$ and $J_0^\t{h}$ describe the effects of the nonmagnetic correlations.
Apart from the term proportional to $\bs\omMn$ in Eq.~\eqref{eq EOM for Mn}, which describes the mean-field precession of the impurity spins around the external magnetic field,
all other contributions in Eqs.~\eqref{eq EOM for Mn}-\eqref{eq EOM for N} can be interpreted analogously.

A similar classification is possible for the source terms of the correlations in Eqs.~\eqref{eq EOM for Q_Y}-\eqref{eq EOM for N_bar}:
Source terms with the upper index $I$ contain inhomogeneous driving terms that only depend on the coherences $Y_{x_1 \0}^{v_1 l_1}$ and the exciton densities $N_{x_1 \K_1}^{l_1 v_1 v_2 l_2}$
and not on carrier-impurity correlations.
The index $II$ denotes homogeneous contributions that cause a precession-type motion of the correlations in the effective fields given by Eqs.~\eqref{eq mean-field frequencies}.
Finally, terms labeled by the index $III$ describe an incoherent driving of the magnetic and nonmagnetic correlations by other carrier-impurity correlations with different wave vectors.

It is noteworthy that, in the absence of an electric field, Eqs.~\eqref{eq full equations of motion} conserve the number of particles as well as the total energy comprised of 
mean-field and correlation contributions, which can be confirmed by a straightforward but lengthy analytical calculation.
This provides an important consistency check of the equations and can be used as a convergence criterion for the numerical implementation.

\subsection{Reduced equations for exciton-bound electron spins}
\label{subsec Reduced equations for exciton-bound electron spins}

An optical excitation with circularly polarized light generates excitons composed of electrons and holes with corresponding electron and hole spins in accordance with the selection rules.
Here, we are dealing with a narrow semiconductor quantum well, where the hh and lh bands are split at the $\Gamma$ point of the Brillouin zone due to the confinement as well as 
strain\cite{Winkler_Spin-Orbit}.
We consider the generation of heavy-hole excitons as they typically constitute the low-energy excitations.
In this case, the hh spins are typically pinned because the precession of a hole spin involves an intermediary occupation of lh states which lie at higher energies. 
Furthermore, for direct transitions between the $-\frac{3}{2}$ and $\frac{3}{2}$ hh states, the corresponding matrix elements in the Hamiltonian given by Eq.~\eqref{eq complete Hamiltonian}
are zero.
As a consequence, if the hh-lh splitting is large enough, hh spins do not take part in the spin dynamics and the initially prepared hole spin does not change.
Therefore, it is sufficient to concentrate only on the dynamics of the exciton-bound electron spins, which can be described by a reduced set of equations of motion.

In the following, we focus on an excitation with $\sigma^-$ polarization, so that heavy-holes with $m_J = -\frac{3}{2}$ and electrons in the spin-up state $\uparrow$ are excited.
Then, it is instructive to consider the dynamical variables
\begin{subequations}
\label{eq dynamical variables summed}
\begin{align}
n_{x_1 K_1} &= \frac{1}{2\pi}\int_0^{2\pi} \!\! d\psi_1 \, \sum_{\sigma} N_{x_1 \K_1}^{\sigma \sigma},
	\an
\mbf s_{x_1 K_1} &= \frac{1}{2\pi}\int_0^{2\pi} \!\! d\psi_1 \, \sum_{\sigma \sigma'} \mbf s_{\sigma \sigma'} N_{x_1 \K_1}^{\sigma \sigma'},
	\an
y_{x_1}^\ud &= \frac{1}{2\pi}\int_0^{2\pi} \!\! d\psi_1 \, Y_{x_1 \0}^\ud,
	\an
q_{\eta l x_1 K_1}^{\phantom{\eta} \ud x_2} &= \frac{1}{2\pi}\int_0^{2\pi} \!\! d\psi_1 \, f_{\eta x_2 x_1}^{\phantom{\eta} \0 \K_1} \int \! dz |u_0(z)|^2 
	\sum_{n n'} S_{n n'}^l 
	\nn
	&\phantom{=\;} \times {Q_Y}_{n x_1 \K_1}^{n' \ud}(z),
	\an
z_{\eta x_1 K_1}^{\phantom{\eta} \ud x_2} &= \frac{1}{2\pi}\int_0^{2\pi} \!\! d\psi_1 \, f_{\eta x_2 x_1}^{\phantom{\eta} \0 \K_1} \int \! dz |u_0(z)|^2 \bar{Y}_{x_1 \K_1}^{\ud}(z),
	\an
Q_{\eta l x_1 K_1}^{\phantom{\eta} \alpha x_2 K_2} &= \frac{1}{4\pi^2}\int_0^{2\pi} \!\! d\psi_1 \int_0^{2\pi} \!\! d\psi_2 \,
	f_{\eta x_1 x_2}^{\phantom{\eta} \K_1 \K_2} \! \int \! dz |u_0(z)|^2  
	\nn 
	&\phantom{=\;} \times \sum_{\substack{\sigma \sigma' \\ n n'}} S_{n n'}^l s_{\sigma \sigma'}^\alpha {Q_N}_{n x_1 \K_1 x_2 \K_2}^{n' \sigma \sigma'}(z),
	\an
Z_{\eta \phantom{\alpha} x_1 K_1}^{\phantom{\eta} \alpha x_2 K_2} &= \frac{1}{4\pi^2}\int_0^{2\pi} \!\! d\psi_1 \int_0^{2\pi} \!\! d\psi_2 \,
	f_{\eta x_1 x_2}^{\phantom{\eta} \K_1 \K_2} \int \! dz |u_0(z)|^2 
	\nn
	&\phantom{=\;} \times \sum_{\sigma \sigma'} s_{\sigma \sigma'}^\alpha \bar{N}_{x_1 \K_1 x_2 \K_2}^{\sigma \sigma'}(z)
\end{align}
\end{subequations}
with $l \in \{1,2,3\}$ and $\alpha \in \{0,1,2,3\}$, where $s_{\sigma_1 \sigma_2}^0 = \delta_{\sigma_1,\sigma_2}$.
We have introduced an average over polar angles $\psi_i$ of the wave vectors $\K_i$, which does not introduce a further approximation in an isotropic system as defined by the 
Hamiltonian in Eq.~\eqref{eq complete Hamiltonian} but significantly reduces the numerical demand.
In Eqs.~\eqref{eq dynamical variables summed}, $n_{x_1 K_1}$ is the occupation density of the excitons with quantum number $x_1$ and modulus of the center of mass momentum 
$K_1$ and $s_{x_1 K_1}$ describes the spin density of exciton-bound electrons.
The interband coherences are described by $y_{x_1}$ and the remaining variables are correlation functions modified by the form factors $f_\eta$ defined in Eq.~\eqref{eq exciton form factors}.

Note that, in order to obtain a closed set of equations for the dynamical variables defined in Eqs.~\eqref{eq dynamical variables summed} starting from Eqs.~\eqref{eq full equations of motion}, 
the source terms $\beta^{III}$, $\bar\beta^{III}$, $b^{III}$, and $\bar b^{III}$ have to be neglected.
However, since these terms contain only sums of correlations with different wave vectors, they can be expected to dephase very fast compared to the remaining source terms.
In previous works on the spin dynamics of conduction band electrons\cite{Cygorek_Comparison-between}, similar terms were shown to be irrelevant by numerical studies.
Furthermore, the optically generated carrier density is typically much lower than the number of impurity atoms in the sample.
This results in a negligible change of the impurity spin over time which is therefore disregarded.

With these assumptions, quantum kinetic equations of motion for the variables defined in Eqs.~\eqref{eq dynamical variables summed} can be derived. 
The results are given in appendix~\ref{ap Quantum kinetic equations of motion with pinned hole spin} where we have introduced the angle-averaged products of form factors
\begin{align}
\label{eq angle-averaged exciton form factors}
&F_{\eta_1 x_1 x_2}^{\eta_2 K_1 K_2} := \frac{1}{4\pi^2} \int_0^{2\pi} \! d\psi_1 \int_0^{2\pi} \! d\psi_2 f_{\eta_1 x_1 x_2}^{\phantom{\eta_1} \K_1 \K_2} 
	\big(f_{\eta_2 x_1 x_2}^{\phantom{\eta_2} \K_1 \K_2}\big)^*
	\nn
	&= 2\pi \! \int_0^{2\pi} \!\!\!\! d\psi \! \int_0^\infty \!\!\!\! dr \! \int_0^\infty \!\!\! dr' \, r r' R_{n_1}(r) R_{n_2}(r) R_{n_1}(r')
	\nn
	&\phantom{=\;} \times R_{n_2}(r') J_{l_1-l_2}\big( \eta_1 K_{12}(\psi) r \big) J_{l_1-l_2}\big( \eta_2 K_{12}(\psi) r' \big)
\end{align}
which contain the influence of the exciton wave function on the spin dynamics.
In the second step, we have used the expansion in Eq.~\eqref{eq exciton form factors} together with the fact that $K_{12} = |\K_1 - \K_2|$ depends only on the 
angle $\psi$ between $\K_1$ and $\K_2$.
For infinite confinement potentials, the influence of the envelope functions defined in Eq.~\eqref{eq envelope function approximation} enters the spin dynamics via the factor
\begin{align}
\label{eq envelope factor}
I &= d\int_{-\frac{d}{2}}^\frac{d}{2} dz |u_0(z)|^4 = \frac{3}{2}.
\end{align}

Note that Eqs.~\eqref{eq EOM for summed variables} also contain second moments of the impurity spin given by
$\langle S^i S^j \rangle = \sum_{n_1 n_2 n_3} S_{n_1 n_2}^i S_{n_2 n_3}^j M_{n_1}^{n_3}$.
Instead of deriving equations of motion for these second moments, we once more exploit the fact that the carrier density is typically much lower than the impurity density,
so that the impurity density matrix is well described by its initial thermal equilibrium value throughout the dynamics\cite{Cygorek_Comparison-between}.

\subsection{Markov limit}
\label{subsec Markov limit}

While the dynamics can in general contain memory effects mediated by carrier-impurity correlations, it is also instructive to consider the Markovian limit of the quantum
kinetic theory, where an infinitesimal memory is assumed.
On the one hand, this allows one to obtain analytical insights into the spin-flip processes described by the theory.
On the other hand, a comparison between quantum kinetic and Markovian results facilitates the identification of true non-Markovian features and allows an estimation of the 
importance of correlations in the system.

To derive the Markov limit, we formally integrate Eqs.~\eqref{eq EOM for summed variables Q1}-\eqref{eq EOM for summed variables Z2} for the correlations.
Afterwards, the resulting integral expressions for the correlations are fed back into Eqs.~\eqref{eq EOM for summed variables n} and \eqref{eq EOM for summed variables s} for the occupation 
densities $n_{x K}$ and the spin densities $\mbf s_{x K}$, respectively.
This yields integro-differential equations for $n_{x K}$ and $\mbf s_{x K}$ alone.
In the Markov limit, the memory integral in these equations is eliminated by assuming that the memory is short so that one can apply the Sokhotsky-Plemelj formula
\begin{align}
\label{eq Sokhotsky-Plemelj}
\int_0^t dt' e^{i\Delta\omega(t'-t)} \stackrel{t \rightarrow \infty}{\longrightarrow} \pi\delta(\Delta\omega) - \frac{i}{\Delta\omega}.
\end{align}
Note that, if a spin precession becomes important, such as in finite magnetic fields, the precession-type motion of carrier and impurity spins as well as of carrier-impurity correlations
have to be treated as fast oscillating contributions that have to be split off in order to identify slowly varying terms that can be drawn out of the memory integral
\cite{Cygorek_Comparison-between}.
This procedure is similar to a rotating-wave description.
The precession frequencies then lead to a modification of $\Delta\omega$ in Eq.~\eqref{eq Sokhotsky-Plemelj} which, in the Markov limit, corresponds to additional energy shifts
that ensure energy conservation during spin-flip processes\cite{Cygorek_Carrier-impurity}.

In the following, we consider a situation where the impurity magnetization as well as the precession vectors are parallel or antiparallel to the external magnetic field.
Then, we can write
\begin{subequations}
\begin{align}
\bs\ome &= \sigma_\t{e}^B \ome \mbf e_B,
	\an
\bs\omh &= \sigma_\t{h}^B \omh \mbf e_B,
	\an
\bs\omMn &= \sigma_\t{Mn}^B \omMn \mbf e_B,
	\an
\langle \mbf S \rangle &= \sigma_S^B \Spar \mbf e_B,
\end{align}
\end{subequations}
where the factors $\sigma_\t{e}^B, \sigma_\t{h}^B, \sigma_\t{Mn}^B, \sigma_S^B \in \{-1,1\}$ determine the direction of the corresponding vector with respect to the direction of the
magnetic field $\mbf e_B$.
It is convenient to choose the variables
\begin{subequations}
\begin{align}
n_{x_1 K_1}^\ud &= \frac{1}{2}n_{x_1 K_1} \pm \mbf s_{x_1 K_1} \cdot \mbf e_B,
	\an
\mbf s_{x_1 K_1}^\perp &= \mbf s_{x_1 K_1} - \big(\mbf s_{x_1 K_1} \cdot \mbf e_B \big) \mbf e_B,
\end{align}
\end{subequations}
which describe the spin-up and spin-down exciton density as well as the perpendicular exciton-bound electron spin density, respectively. 
For these variables, the Markovian equations of motion are:
\begin{widetext}
\begin{subequations}
\label{eq Markov equations}
\begin{align}
\label{eq Markov equation n}
\ddt n_{x_1 K_1}^\ud &= 
	\Gamma_{\mbf E}^\ud +  \frac{\pi I \NMn}{\hbar^2 V^2} \sum_{x K} \bigg\{ \delta\big(\omega_{x K} \! - \! \omega_{x_1 K_1}\big) \Big(n_{x K}^\ud - n_{x_1 K_1}^\ud\Big)
	\! \Big[ \big( \Jsd^2 b^\parallel \pm 2\Jsd J_0^\t{e} b^0 + 2{J_0^\t{e}}^2 \big) F_{\etah x x_1}^{\etah K K_1}
	\nn
	&\phantom{=\;} + \big( \Jpd^2 b^\parallel - 2\Jpd J_0^\t{h} b^0 + 2{J_0^\t{h}}^2 \big) F_{\etae x x_1}^{\etae K K_1}
	+ \big( 4J_0^\t{e}J_0^\t{h} - 2\Jpd J_0^\t{e} b^0 \pm 2\Jsd J_0^\t{h} b^0 \mp 2\Jsd\Jpd b^\parallel \big) F_{-\etah x x_1}^{\phantom{-} \etae K K_1} \Big] 
	\nn
	&\phantom{=\;} + \delta\big(\omega_{x K} \! - \! \big(\omega_{x_1 K_1} \! \pm \! (\sigma_\t{e}^B\ome \! - \! \sigma_\t{Mn}^B\omMn)\big)\big) \Jsd^2 
	F_{x x_1}^{K K_1} \Big(b^\pm n_{x K}^\du - b^\mp n_{x_1 K_1}^\ud\Big) \bigg\},
	\an
\label{eq Markov equation s}
\ddt \mbf s_{x_1 K_1}^\perp &= 
	\bs\Gamma_{\mbf E}^\perp + \frac{\pi I \NMn}{\hbar^2 V^2} \sum_{x K} \bigg\{ 
	\delta\big(\omega_{x K} \! - \! \omega_{x_1 K_1}\big) \Big(\mbf s_{x K}^\perp - \mbf s_{x_1 K_1}^\perp\Big)
	\Big[ \big( 2{J_0^\t{e}}^2 - \Jsd^2 b^\parallel \big) F_{\etah x x_1}^{\etah K K_1} 
	+ \big( \Jpd^2 b^\parallel + 2{J_0^\t{h}}^2 - \Jpd J_0^\t{h} b^0 \big)
	\nn
	&\phantom{=\;} \times F_{\etae x x_1}^{\etae K K_1} - \big( 2\Jpd J_0^\t{e} b^0  + \Jpd J_0^\t{h} b^0 - 4J_0^\t{e} J_0^\t{h} \big) F_{-\etah x x_1}^{\phantom{-} \etae K K_1} \Big]
	- \bigg[ \frac{b^-}{2} \delta\big(\omega_{x K} \! - \! \big(\omega_{x_1 K_1} \! + \! (\sigma_\t{e}^B\ome \! - \! \sigma_\t{Mn}^B\omMn)\big)\big) 
	\nn
	&\phantom{=\;} + \frac{b^+}{2} \delta\big(\omega_{x K} \! - \! \big(\omega_{x_1 K_1} \! - \! (\sigma_\t{e}^B\ome \! - \! \sigma_\t{Mn}^B\omMn)\big)\big) 
	+ 2 b^\parallel \delta\big(\omega_{x K} \! - \! \omega_{x_1 K_1}\big) \bigg] \Jsd^2 F_{\etah x x_1}^{\etah K K_1} \mbf s_{x_1 K_1}^\perp \bigg\}
	\nn
	&\phantom{=\;} + \Big( \bs\ome \times \mbf s_{x_1 K_1}^\perp \Big) \Bigg\{ 1 + \frac{1}{\ome}\frac{I \NMn}{\hbar^2 V^2} \sum_{x K} \bigg[ \frac{\Jsd}{\omega_{x K} \! - \! \omega_{x_1 K_1}} 
	\Big(\big( 2\Jpd b^\parallel - 2J_0^\t{h} b^0 \big) F_{-\etah x x_1}^{\phantom{-} \etae K K_1} - 2 J_0^\t{e} b^0 F_{\etah x x_1}^{\etah K K_1} \Big)
	\nn
	&\phantom{=\;} + \bigg( \frac{b^+}{\omega_{x K} \! - \! \big(\omega_{x_1 K_1} \! - \! (\sigma_\t{e}^B\ome \! - \! \sigma_\t{Mn}^B\omMn)\big)} 
	- \frac{b^-}{\omega_{x K} \! - \! \big(\omega_{x_1 K_1} \! + \! (\sigma_\t{e}^B\ome \! - \! \sigma_\t{Mn}^B\omMn)\big)} \bigg)
	\frac{1}{2}\Jsd^2 F_{\etah x x_1}^{\etah K K_1} \bigg]\Bigg\}.
\end{align}
\end{subequations}
\end{widetext}

\begin{figure*}[t!]
	\includegraphics{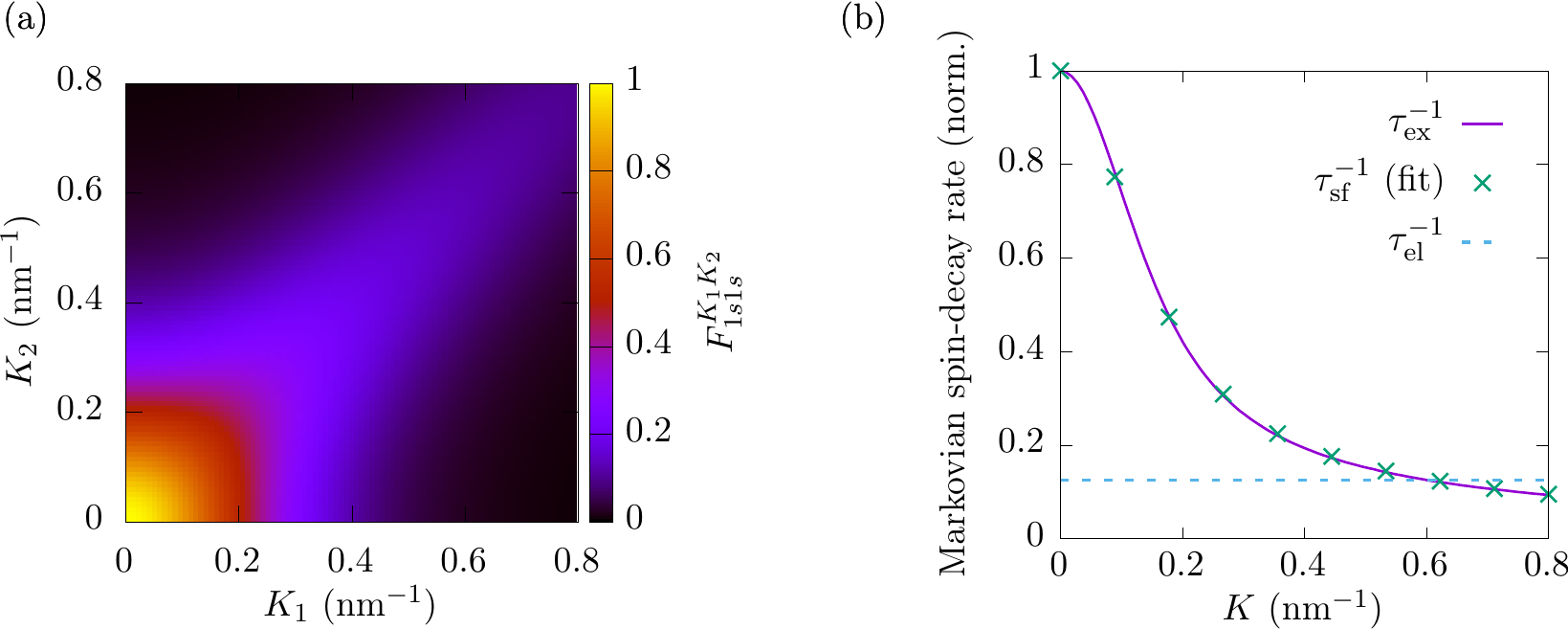}
	\caption{(Color online) (a) Angle-averaged form-factor product $F_{\etah 1s 1s}^{\etah K_1 K_2}$ for the exciton ground state (1s) as a function of the center of mass momentum $K$.
	(b) $K$-dependence of the Markovian spin-decay rate for excitons ($\tau_\t{ex}^{-1}$), which follows the diagonal of plot (a), is compared to the constant rate found for 
	quasi-free electrons\cite{Cygorek_Carrier-impurity} ($\tau_\t{el}^{-1}$).
	Both rates are normalized with respect to the exciton spin-decay rate for $K = 0$ and are calculated without external magnetic field.
	Additionally, Eq.~\eqref{eq Bastard spin-flip rate} is used to fit $\tau_\t{sf}^{-1}$ to the exciton spin-decay rate of our model.
	\label{fig matrix elements}}
\end{figure*}

In the above equations, the shorthand notation $b^\pm := \frac{1}{2}\big(\langle \mbf S^2 - (\mbf S \cdot \mbf e_B)^2\rangle \pm \sigma_S^B \Spar\big)$, 
$b^\parallel := \frac{1}{2} \langle (\mbf S \cdot \mbf e_B)^2\rangle$, and $b^0 := \sigma_S^B \Spar$ has been used for the second moments of the Mn spin.
Here, we model the optical excitation by the generation rates $\Gamma_{\mbf E}^\ud$ and $\bs\Gamma_{\mbf E}^\perp$ for the spin-up and spin-down occupations and the perpendicular 
spin component, respectively.

In Eq.~\eqref{eq Markov equation n}, the term proportional to $\big(n_{x K}^\ud - n_{x_1 K_1}^\ud\big)$ describes processes conserving the exciton spin, whereas the term 
proportional to $\big(b^\pm n_{x K}^\du - b^\mp n_{x_1 K_1}^\ud\big)$ is responsible for the spin-flip scattering of excitons.
The delta functions ensure conservation of energy.
Similarly, the terms proportional to $\big(\mbf s_{x K}^\perp - \mbf s_{x_1 K_1}^\perp\big)$ in Eq.~\eqref{eq Markov equation s} can be interpreted as exciton-spin conserving
contributions, whereas the prefactors of $s_{x_1 K_1}^\perp$ are responsible for a decay of the perpendicular spin component.
Finally, the cross product $\bs\ome \times \mbf s^\perp_{x_1 K_1}$ describes the mean-field precession around $\bs\ome$ which is renormalized by terms resulting from the imaginary
part of the memory integral given by Eq.~\eqref{eq Sokhotsky-Plemelj}.

The exciton spin-conserving parts of Eqs.~\eqref{eq Markov equations} lead to a redistribution within a given energy shell as well as to transitions between excitonic states with different
quantum numbers, as can be seen from the argument of the corresponding delta functions.
In situations where spin-orbit coupling and thus a D'yakonov-Perel'-type spin dephasing is important, these terms give rise to an additional momentum scattering and thereby indirectly
influence the spin dynamics.
However, spin-orbit coupling is typically of minor importance for the spin dynamics in DMS compared with the carrier-impurity interaction\cite{Ungar_Ultrafast-spin}.
In an isotropic system as considered here, the exciton spin-conserving parts of Eqs.~\eqref{eq Markov equations} do not influence the spin dynamics.
Since the magnetic coupling constant $\Jpd$ for the valence band as well as the non-magnetic coupling constants $J_0^\t{e}$ and $J_0^\t{h}$ only enter these terms, the nonmagnetic 
interactions and the $pd$ interaction do not affect the spin dynamics on the Markovian level.

For spin-flip scattering processes, an exciton with a given spin an energy $\hbar\omega_{x_1 K_1}$ is scattered to a state with opposite spin and energy $\hbar\omega_{x K}$.
The appearance of the energy shift $\pm \hbar (\sigma_\t{e}^B\ome - \sigma_\t{Mn}^B\omMn)$ in the corresponding delta function in Eq.~\eqref{eq Markov equation n} 
can be understood as follows:
A flip of the exciton-bound electron spin requires or releases a magnetic energy $\hbar\sigma_\t{e}^B\ome$.
But since a flip of a carrier spin also involves the flop of an impurity spin in the opposite direction, the corresponding change in magnetic energy of the impurity spin
$\hbar\sigma_\t{Mn}^B\omMn$ has to be accounted for to ensure conservation of energy.

An interesting limiting case can be worked out for zero external magnetic field, vanishing impurity magnetization, and optical excitation resonant with the $1s$ exciton state:
Then, Eqs.~\eqref{eq Markov equations} can be condensed into the simple rate equation
\begin{align}
\ddt \mbf s_{1s K_1} &= - \tau^{-1}_{1s K_1} \mbf s_{1s K_1},
\end{align}
where the spin-decay rate is given by
\begin{align}
\label{eq Markov spin-transfer rate}
\tau^{-1}_{1s K_1} &= \frac{35}{12} \frac{\NMn I \Jsd^2 M}{\hbar^3 d V} F_{\etah 1s 1s}^{\etah K_1 K_1}
\end{align}
and $d$ denotes the width of the DMS quantum well.
In contrast to the quasi-free electron case, where the spin-decay rate is constant in a quantum well\cite{Kossut_On-the, Wu_Spin-dynamics}, the decay rate for excitons
explicitly depends on $K$, which is consistent with previous findings in the literature\cite{Bastard_Spin-flip}.

\section{Results}
\label{sec Results}

We now apply our quantum kinetic theory to the exciton spin dynamics for vanishing external magnetic field and impurity magnetization after an ultrashort laser pulse resonant with the
exciton ground state and compare the results with the corresponding Markovian calculations.
In order to do so, it is necessary to first calculate the exciton wave functions and the resulting form-factor products $F_{\eta_1 x_1 K_1}^{\eta_2 x_2 K_2}$.

\subsection{Exciton form factors}
\label{subsec Exciton form factors}

In order to calculate the exciton form factors, we first decompose the exciton wave function according to Eq.~\eqref{eq decomposition of wave function} and then numerically solve the Coulomb 
eigenvalue problem given by Eq.~\eqref{eq Exciton problem} for the radial part using a finite-difference method, which yields the exciton energies as well as the wave functions.
From the exciton wave functions, the form-factor products defined in Eq.~\eqref{eq angle-averaged exciton form factors} are calculated.
The steps and cut-offs in the real-space discretization have been adjusted to ensure convergence.

The results for the form-factor product $F_{\etah 1s 1s}^{\etah K_1 K_2}$ relevant for spin-slip scattering on the $1s$ exciton parabola can be found in Fig.~\ref{fig matrix elements}(a)
as a function of wave numbers $K_1$ and $K_2$ using the parameters for Cd$_{1-x}$Mn$_x$Te listed in Tab.~\ref{tab material parameters}.
It can be seen that $F_{\etah 1s 1s}^{\etah K_1 K_2}$ is symmetric with respect to the bisectrix and decreases continuously with increasing wave number.
In Fig.~\ref{fig matrix elements}(b), we present the spin-decay rate in the Markov limit according to Eq.~\eqref{eq Markov spin-transfer rate} which follows the diagonal values 
$F_{\etah 1s 1s}^{\etah K_1 K_1}$.
To compare the resulting rate to the quasi-free electron case, we also plot the spin-decay rate from Ref.~\onlinecite{Cygorek_Influence-of} for electrons and normalize 
both results to the exciton spin-decay rate for $K = 0$.
The spin-decay rate for excitons at $K = 0$ is about 8 times faster than the electron spin-decay rate, which is due to the much larger exciton mass.
Furthermore, the exciton spin-decay rate strongly depends on $K$ and can even be smaller than the constant electron spin-decay rate for large wave numbers.

The fact that the spin-decay rate for excitons depends on $K$ has already been pointed out in Ref.~\onlinecite{Bastard_Spin-flip}. 
There, an exponential ansatz with a variational parameter for the radial part of the exciton wave function leads to the decay rate\cite{Bastard_Spin-flip}
\begin{align}
\label{eq Bastard spin-flip rate}
\frac{1}{\tau_\t{sf}} (K) &= \frac{1}{\tau_\t{sf}}(0) \, \phi(\alpha^2 K^2),
\end{align}
where the constant $\alpha$ contains the parameters of the model and the function $\phi$ is given by\cite{Bastard_Spin-flip}
\begin{align}
\phi(x) &= \frac{1}{2} \big(1+(1+x)\big) \big(1+2x\big)^{-\frac{5}{2}}.
\end{align}
To compare this result to our calculations, we fit the constant $\alpha$ in Eq.~\eqref{eq Bastard spin-flip rate} to our data obtained from Eq.~\eqref{eq Markov spin-transfer rate}
and plot the result in Fig.~\ref{fig matrix elements}(b).
It can be seen that the predictions of Ref.~\onlinecite{Bastard_Spin-flip} agree with the Markovian limit of our quantum kinetic theory.

\subsection{Spin dynamics}
\label{subsec Spin dynamics}

Having obtained the exciton form factors, we can now calculate the spin dynamics according to the quantum kinetic Eqs.~\eqref{eq EOM for summed variables}.
To address the question of the importance of quantum kinetic effects in the exciton spin dynamics, we also present numerical solutions of the Markovian Eqs.~\eqref{eq Markov equations}.
Furthermore, we study the influence of nonmagnetic scattering as well as the magnetic $pd$ coupling.

For the numerical implementation, we use a forth-order Runge-Kutta algorithm to solve the differential equations in the time domain and discretize the $K$ space up to a cut-off 
energy of a few tens of meV.
This is done in the quasi-continuous limit $\sum_K \to \int dK D^{2d}(K)$ using the two-dimensional density of states $D^{2d}(K) = \frac{A}{2\pi}K$ for a quantum well with area $A$.
For all calculations, we have checked that the number of excitons in the system as well as the total energy remain constant after the pulse.

\begin{figure*}[ht!]
	\includegraphics{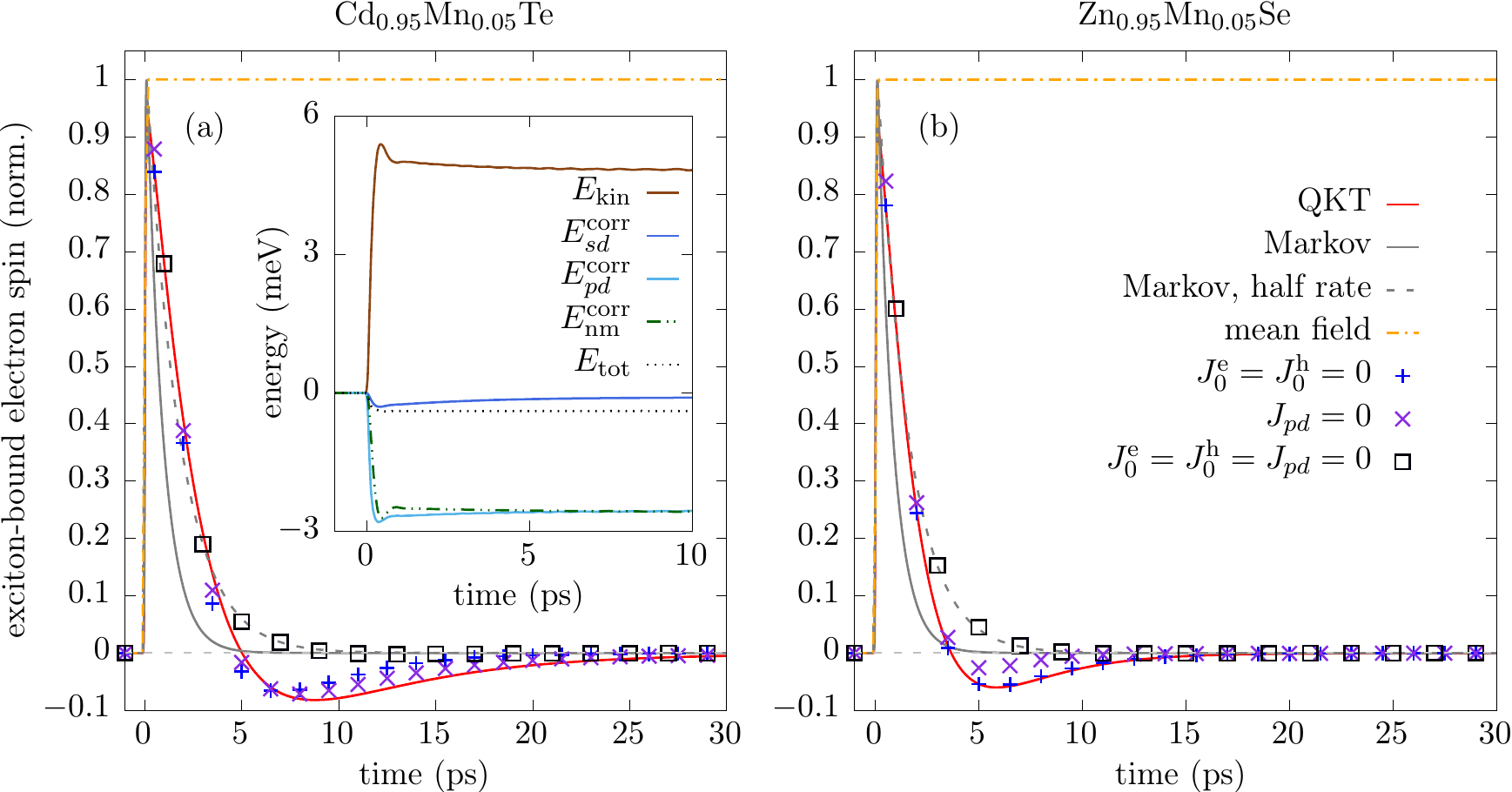}
	\caption{(Color online) Spin dynamics of the exciton-bound electron in a $10\,$nm quantum well using the parameters of (a) Cd$_{0.95}$Mn$_{0.05}$Te and (b) Zn$_{0.95}$Mn$_{0.05}$Se
	for vanishing external magnetic field after an optical excitation with a circularly polarized Gaussian laser beam resonant to the exciton ground state.
	The time axis is chosen such that the pulse maximum lies at $t = 0\,$ps with a width (FWHM) of $100\,$fs.
	For each material, we show the quantum kinetic results based on Eqs.~\eqref{eq EOM for summed variables} (QKT) as well as Markovian calculations using 
	Eqs.~\eqref{eq Markov equations} (Markov), a spin decay with half the Markovian rate (Markov, half rate), and the results of a calculation where all correlations are 
	neglected (mean field).
	Additionally, we plot the spin dynamics in the artificial situation where different coupling constants $J_0^\t{e}$, $J_0^\t{h}$ and/or $\Jpd$ are set to zero.
	All results are normalized with respect to the mean-field spin polarization for long times.
	The inset in figure (a) shows the kinetic energy ($E_\t{kin}$), the magnetic $sd/pd$ correlation energies ($E_{sd/pd}^\t{corr}$), the nonmagnetic correlation energy ($E_\t{nm}^\t{corr}$), 
	and the total energy ($E_\t{tot}$) normalized with respect to the exciton density after the pulse.
	\label{fig exciton spin dynamics}}
\end{figure*}

We limit our study to the exciton ground state and treat the optical excitation in a rotating-wave approximation.
As discussed in section \ref{subsec Reduced equations for exciton-bound electron spins}, we focus on a situation where the hh spins are pinned and do not take part in the dynamics.
Thus, our main quantity of interest is the time evolution of the spin of the exciton-bound electron.
In all cases, the optical excitation is modeled by a circularly polarized Gaussian laser beam with a width (FWHM) of $100\,$fs centered at $t = 0\,$ps resonant to the exciton ground state
and we consider a quantum well with width $d = 10\,$nm.
We calculate the time evolution of the exciton spin for two different materials, namely Cd$_{1-x}$Mn$_x$Te [Fig.~\ref{fig exciton spin dynamics}(a)] as well as Zn$_{1-x}$Mn$_x$Se 
[Fig.~\ref{fig exciton spin dynamics}(b)] with impurity concentration $x = 5\,\%$.
The relevant parameters for these two materials, which are both of zinc blende crystal structure\cite{Furdyna_Diluted-magnetic}, are collected in Tab.~\ref{tab material parameters}.

\begin{table}[ht!]
	\begin{tabular}{ccc}
	\hline\hline
	parameter & Cd$_{1-x}$Mn$_x$Te & Zn$_{1-x}$Mn$_x$Se\\
	\hline
	$a\;(\t{nm})$\cite{Furdyna_Diluted-magnetic} & $0.648$ & $0.567$\\
	$m_\t{e}/m_0$\cite{Astakhov_Binding-energy, Triboulet_CdTe-and} & $0.1$ & $0.15$ \\
	$m_\t{hh}/m_0$\cite{Astakhov_Binding-energy, Triboulet_CdTe-and} & $0.7$ & $0.8$ \\
	$\Jsd\;(\t{meV}\,\t{nm}^3)$\cite{Furdyna_Diluted-magnetic} & $-15$ & $-12$ \\
	$\Jpd\;(\t{meV}\,\t{nm}^3)$\cite{Furdyna_Diluted-magnetic} & $60$ & $50$ \\
	$J_0^\t{e}\;(\t{meV}\,\t{nm}^3)$\cite{Furdyna_Diluted-magnetic} & $110$ & $22$ \\
	$J_0^\t{h}\;(\t{meV}\,\t{nm}^3)$\cite{Furdyna_Diluted-magnetic} & $7$ & $0$ \\
	$\epsilon$\cite{Strzalkowski_Dielectric-constant} & $10$ & $9$ \\
	\hline\hline
	\end{tabular}
	\caption{Selected material parameters of Cd$_{1-x}$Mn$_x$Te and Zn$_{1-x}$Mn$_x$Se.
	The coupling constant is chosen such that it is consistent with the band offsets at a CdTe/Cd$_{1-x}$Mn$_x$Te and ZnSe/Zn$_{1-x}$Mn$_x$Se interface, 
	respectively\cite{Cygorek_Influence-of}. The cubic lattice constant is given by $a$ and $m_0$ denotes the free electron mass.
	\label{tab material parameters}}
\end{table}

The mean-field results displayed in Fig.~\ref{fig exciton spin dynamics} show no spin decay because the time evolution of the exciton density matrix 
[c.f. Eq.~\eqref{eq EOM for summed variables s}] after the optical excitation in the absence of a magnetic field is governed by the magnetic and nonmagnetic correlations, 
which are neglected in the mean-field approximation.
If the correlations are treated on a Markovian level, the spin decays exponentially with the spin-decay rate $\tau_{1s K_1}^{-1}$ defined in Eq.~\eqref{eq Markov spin-transfer rate}.
The spin dynamics in Cd$_{0.95}$Mn$_{0.05}$Te is slower than in Zn$_{0.95}$Mn$_{0.05}$Se, which is mainly due to the larger exciton mass in ZnSe.

However, the full quantum kinetic spin dynamics in both materials is clearly non-monotonic and shows a pronounced overshoot after approximately $5\,$ps of about $10\,\%$
of the spin polarization immediately after the pulse in the situation depicted in Fig.~\ref{fig exciton spin dynamics}(a).
Furthermore, for the first few picoseconds, the quantum kinetic result is actually closer to the results of a calculation using only half the Markovian spin-decay rate.
The spin overshoot in Fig.~\ref{fig exciton spin dynamics} is absent if the nonmagnetic impurity scattering of electrons and holes in the DMS as well as the $pd$ exchange interaction 
are neglected, as suggested by a calculation with $J_0^\t{e} = J_0^\t{h} = \Jpd = 0$ (c.f. black boxes in Fig.~\ref{fig exciton spin dynamics}).
Without these contributions, the time evolution of the spin virtually coincides with an exponential decay with half the Markovian spin-decay rate.
Note that the nonmagnetic scattering as well as the $pd$ interaction do not influence the spin dynamics on the Markovian level, as follows from Eqs.~\eqref{eq Markov equations}.

Interestingly, the role of non-magnetic impurity scattering is here opposite to what has been found for the electron spin dynamics in the band continuum\cite{Cygorek_Influence-of}:
While for excitonic excitations this scattering enhances the overshoot, for above band-gap excitations it typically almost completely suppresses the non-monotonic time dependence
of the electron spin polarization.

\begin{figure}
	\includegraphics[scale=0.35]{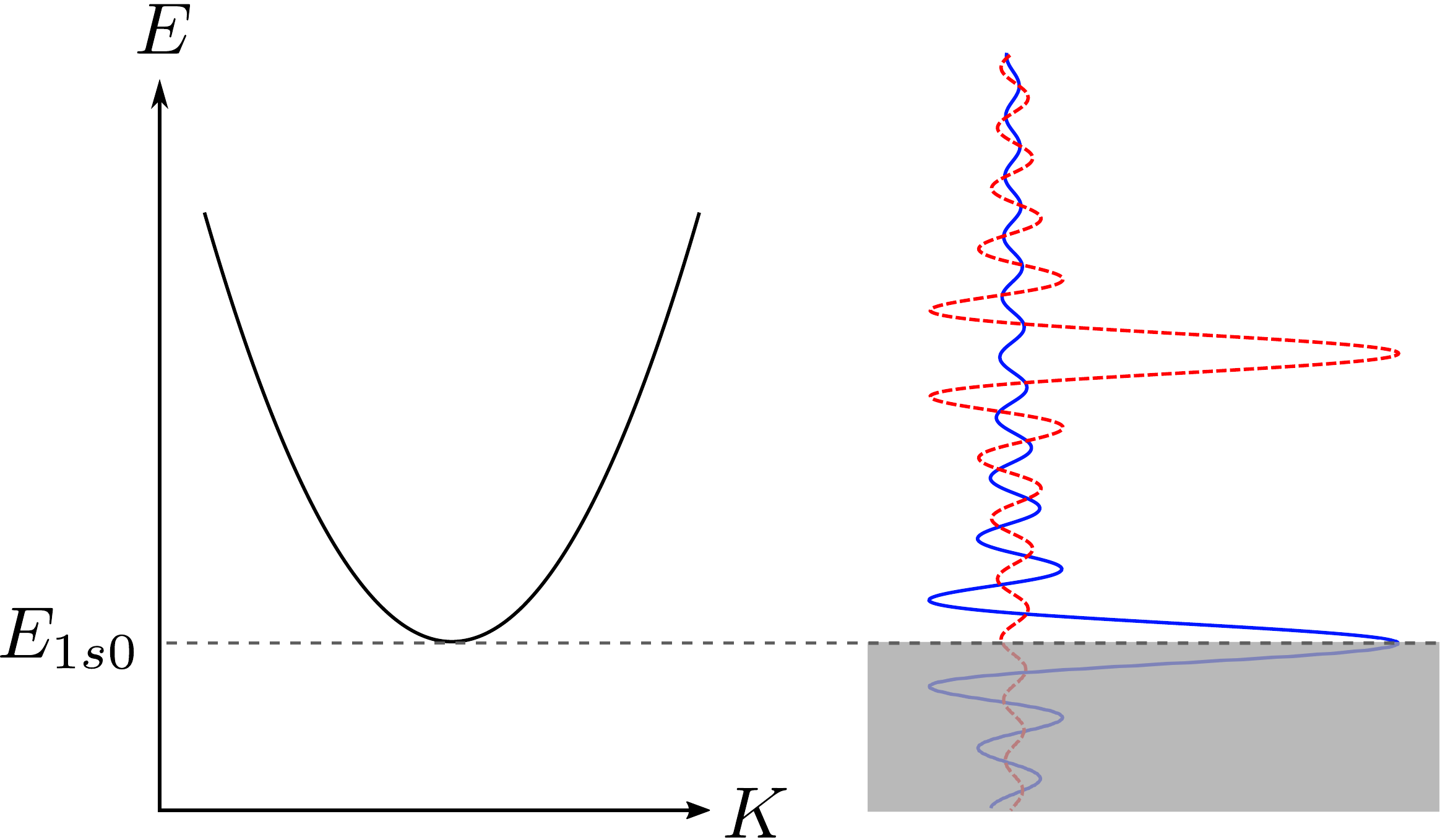}
	\caption{(Color online) Sketch of the $1s$ exciton parabola together with the real part of a typical memory kernel $\propto \frac{\sin[(E-E_1)t/\hbar]}{(E-E_1)/\hbar}$ 
	[c.f. l.h.s. of Eq.~\eqref{eq Sokhotsky-Plemelj}] for a fixed time $t$ with $E_1 = E_{1s 0}$ (blue solid line) and $E_1 > E_{1s 0}$ (red dashed line).
	For excitons optically generated at $K \approx 0$, the memory is effectively cut in half.
	\label{fig sinc}}
\end{figure}

The deviations from the Markovian limit can be traced back to the optical excitation at the bottom of the exciton parabola ($K \approx 0$):
While the memory kernel in the Markovian limit given by Eq.~\eqref{eq Sokhotsky-Plemelj} contracts to a delta function in energy space, for finite times the energy-time uncertainty relation
leads to a finite spectral width as sketched in Fig.~\ref{fig sinc}.
In a quantum well, the spectral density of states is constant but vanishes below the vertex of the exciton parabola, resulting in a cut-off of the memory kernel in energy
space\cite{Cygorek_Non-Markovian}.
At $K = 0$, the integral over the memory kernel yields therefore only half the value predicted by the Markovian assumption of a delta-like memory.
This translates into a reduction of the effective spin-decay rate by a factor of~$\frac{1}{2}$.

The influence of nonmagnetic impurity scattering manifests itself in a redistribution of center of mass momenta on the $1s$ exciton parabola.
As a result, states further away from $K = 0$ are populated.
For these states, the cut-off of the memory integral due to the density of states is shifted correspondingly and oscillates with time, which causes the overshoots in the 
dynamics of the spin polarization in Fig.~\ref{fig exciton spin dynamics}.
It is noteworthy that, even if the heavy-hole spins are pinned throughout the dynamics, the magnetic $pd$ exchange interaction can still influence the dynamics of the exciton-bound
electron spin since it allows for spin-conserving scattering of exciton-bound holes at magnetic impurities.
In this sense, the magnetic $pd$ interaction has a similar effect as nonmagnetic scattering.
This can be seen from the results depicted in Fig.~\ref{fig exciton spin dynamics} where either the nonmagnetic or the $pd$ interactions are switched off.
In the case of Cd$_{0.95}$Mn$_{0.05}$Te, where $\Jpd \approx \frac{1}{2} J_0^\t{e}$, both interactions are of similar importance.
However, for Zn$_{0.95}$Mn$_{0.05}$Se, where $\Jpd \approx 2 J_0^\t{e}$, the magnetic $pd$ interaction dominates the spin dynamics and nonmagnetic impurity scattering is 
almost negligible.

The fact that the $pd$ interaction and the nonmagnetic impurity scattering facilitate a redistribution of center of mass momenta can be seen from the inset of 
Fig.~\ref{fig exciton spin dynamics}(a), which shows the time evolution of the kinetic energy as well as correlation energies.
A significant increase in kinetic energy of about $5\,$meV per exciton is found, which is mainly provided by a build-up of correlation energies due to nonmagnetic scattering 
and due to the $pd$ interaction.
The inset in Fig.~\ref{fig exciton spin dynamics}(a) also shows that the total exciton energy is indeed conserved after the pulse and obtains a small negative value
with respect to the mean-field energy of a $1s$ exciton at $K = 0$.
This is possible because carrier-impurity correlations are built up already during the finite width of the pulse.
In the case of Cd$_{0.95}$Mn$_{0.05}$Te the magnetic $pd$ interaction as well as the nonmagnetic impurity scattering lead to similar correlations energies, which is consistent
with their comparable influence on the time evolution of the spin as depicted in the main panel of Fig.~\ref{fig exciton spin dynamics}(a).

\section{Conclusion}
\label{sec Conclusion}

We have derived quantum kinetic equations for density matrices in the exciton representation that describe the time evolution of the exciton spin in laser-driven DMS in the 
presence of an external magnetic field.
Our theory takes into account contributions up to second order in the generating laser field and explicitly keeps correlations between the carrier and the impurity subsystem.
The model not only includes the magnetic $sp$-$d$ interaction between electrons, holes, and Mn atoms, but also accounts for elastic nonmagnetic scattering at the impurities.
This makes our theory a widely applicable tool to study the ultrafast spin dynamics in DMS beyond the single-particle Born-Markov picture.
Furthermore, we have shown how rate equations can be straightforwardly extracted from our quantum kinetic theory by using the Markov approximation to eliminate the correlations.
This approach allows us to obtain spin-flip scattering rates for situations where the spin polarization is oriented parallel or perpendicular with respect to the external magnetic field.
In contrast to the situation of quasi-free conduction band electrons studied in Ref.~\onlinecite{Cygorek_Influence-of}, for excitons it is found that the Markovian spin-decay rate
strongly depends on the wave vector via a form factor reflecting the shape of the exciton wave function.

A numerical solution of the quantum kinetic equations including exciton-impurity correlations in the absence of a magnetic field and for vanishing impurity magnetization 
reveals strong deviations from the Markovian predictions in the form of an overshoot of the spin polarization as well as a slower initial decay with about half of the Markovian rate.
Accounting for nonmagnetic impurity interaction as well as the $pd$ interaction in the valence band was found to have an essential impact on the spin polarization since the
overshoot is only seen in calculations that include these interactions.
In contrast, non-monotonic behavior in the spin dynamics of conduction band electrons is strongly suppressed by nonmagnetic impurity scattering\cite{Cygorek_Influence-of}.

In Ref.~\onlinecite{BenCheikh_Electron-spin}, where results for spin-decay rates in DMS measured by different groups have been compared, it was found that the experimentally obtained
rates for vanishing magnetic field are consistently about a factor of $5$ larger than the value expected from Fermi's golden rule for conduction band electrons.
A possible explanation for this deviation is that excitons instead of quasi-free electrons have to be considered.
Substituting the exciton mass for the electron mass in Fermi's golden rule leads to an approximately 8 times larger spin-decay rate.
However, in this article, we have found that non-Markovian effects lead to a spin decay on a time scale corresponding to about half the Markovian rate.
Thus, our theory predicts that the spin-decay rate measurable in ultrafast optical experiments is about 4 times larger than predicted by a Markovian model using quasi-free carriers
and is therefore close to the findings of experiments.

\section*{Acknowledgments}
\label{sec Acknowledgments}

We gratefully acknowledge the financial support of the Deutsche Forschungsgemeinschaft (DFG) through Grant No. AX17/10-1.

\appendix
\section{Source terms for the correlations}
\label{ap Source terms for the correlations}

The source terms for the correlations in Eqs.~\eqref{eq full equations of motion} are:

\begin{widetext}
\begin{subequations}
\label{eq source terms for the correlations}
\begin{align}
&{\beta_{n_1 x_1 \K_1}^{n_2 v_1 l_1}(z)}^{I} 
	= \sum_{n x} M_{n_1 n}(z) \mbf S_{n_2 n} \cdot \Big( \Jsd d |u_0^\t{e}(z)|^2 \sum_{l} \mbf s^\t{e}_{l_1 l} f_{-\etah x_1 x}^{\phantom{-\etah} \K_1 \0} Y_{x \0}^{v_1 l}
	+ \Jpd d |u_0^\t{h}(z)|^2 \sum_{v} \mbf s^\t{h}_{v_1 v} f_{\etae x_1 x}^{\phantom{\etae} \K_1 \0} Y_{x \0}^{v l_1} \Big)
	\nn
	&\phantom{{\beta_{n_1 x_1 \K_1}^{n_2 v_1 l_1}(z)}^{I}=\,} + \sum_x M_{n_1}^{n_2}(z) \Big( J_0^\t{e} d |u_0^\t{e}(z)|^2 f_{-\etah x_1 x}^{\phantom{-\etah} \K_1 \0}
	+ J_0^\t{h} d |u_0^\t{h}(z)|^2 f_{\etae x_1 x}^{\phantom{\etae} \K_1 \0} \Big) Y_{x \0}^{v_1 l_1},
	\an
&{\beta_{n_1 x_1 \K_1}^{n_2 v_1 l_1}(z)}^{II}
	= \sum_{l} \hbar\bs\ome \cdot \mbf s^\t{e}_{l_1 l} {Q_Y}_{n_1 x_1 \K_1}^{n_2 v_1 l}(z)
	+ \sum_{v} \hbar\bs\omh \cdot \mbf s^\t{h}_{v_1 v} {Q_Y}_{n_1 x_1 \K_1}^{n_2 v l_1}(z)
	\nn
	&\phantom{{\beta_{n_1 x_1 \K_1}^{n_2 v_1 l_1}(z)}^{II}=\,} + \sum_{n} \hbar\bs\omMn \cdot 
	\Big( \mbf S_{n_2 n} {Q_Y}_{n_1 x_1 \K_1}^{n v_1 l_1}(z) - \mbf S_{n n_1} {Q_Y}_{n x_1 \K_1}^{n_2 v_1 l_1}(z) \Big),
	\an
&{\beta_{n_1 x_1 \K_1}^{n_2 v_1 l_1}(z)}^{III} 
	= \frac{1}{V} \sum_{n x \K} \mbf S_{n_2 n} \cdot \Big( \Jsd d |u_0^\t{e}(z)|^2 \sum_{l} \mbf s^\t{e}_{l_1 l} f_{-\etah x_1 x}^{\phantom{-\etah} \K_1 \K} {Q_Y}_{n_1 x \K}^{n v_1 l}(z)
	+ \Jpd d |u_0^\t{h}(z)|^2 \sum_{v} \mbf s^\t{h}_{v_1 v} f_{\etae x_1 x}^{\phantom{\etae} \K_1 \K} {Q_Y}_{n_1 x \K}^{n v l_1}(z) \Big)
	\nn
	&\phantom{{\beta_{n_1 x_1 \K_1}^{n_2 v_1 l_1}(z)}^{III}=\,} + \frac{1}{V} \sum_{x \K} 
	\Big( J_0^\t{e} d |u_0^\t{e}(z)|^2 f_{-\etah x_1 x}^{\phantom{-\etah} \K_1 \K} 
	+ J_0^\t{h} d |u_0^\t{h}(z)|^2 f_{\etae x_1 x}^{\phantom{\etae} \K_1 \K} \Big) {Q_Y}_{n_1 x \K}^{n_2 v_1 l_1}(z),
	\an
&{\bar{\beta}_{x_1 \K_1}^{v_1 l_1}(z)}^{I} 
	= \sum_{n n' x} M_{n}^{n'}(z) \mbf S_{n n'} \cdot \Big( \Jsd d |u_0^\t{e}(z)|^2 \sum_{l} \mbf s^\t{e}_{l_1 l} f_{-\etah x_1 x}^{\phantom{-\etah} \K_1 \0} Y_{x \0}^{v_1 l}
	+ \Jpd d |u_0^\t{h}(z)|^2 \sum_{v} \mbf s^\t{h}_{v_1 v} f_{\etae x_1 x}^{\phantom{\etae} \K_1 \0} Y_{x \0}^{v l_1} \Big)
	\nn
	&\phantom{{\bar{\beta}_{x_1 \K_1}^{v_1 l_1}(z)}^{I}=\,} + \sum_x 
	\Big( J_0^\t{e} d |u_0^\t{e}(z)|^2 f_{-\etah x_1 x}^{\phantom{-\etah} \K_1 \0} 
	+ J_0^\t{h} d |u_0^\t{h}(z)|^2 f_{\etae x_1 x}^{\phantom{\etae} \K_1 \0} \Big) Y_{x \0}^{v_1 l_1},
	\an
&{\bar{\beta}_{x_1 \K_1}^{v_1 l_1}(z)}^{II} 
	= \sum_{l} \hbar\bs\ome \cdot \mbf s^\t{e}_{l_1 l} \bar{Y}_{x_1 \K_1}^{v_1 l}(z)
	+ \sum_{v} \hbar\bs\omh \cdot \mbf s^\t{h}_{v_1 v} \bar{Y}_{x_1 \K_1}^{v l_1}(z),
	\an
&{\bar{\beta}_{x_1 \K_1}^{v_1 l_1}(z)}^{III}
	= \frac{1}{V} \sum_{\substack{n n' \\ x \K}} \mbf S_{n n'} \cdot 
	\Big( \Jsd d |u_0^\t{e}(z)|^2 \sum_{l} \mbf s^\t{e}_{l_1 l} f_{-\etah x_1 x}^{\phantom{-\etah} \K_1 \K} {Q_Y}_{n x \K}^{n' v_1 l}(z)
	+ \Jpd d |u_0^\t{h}(z)|^2 \sum_{v} \mbf s^\t{h}_{v_1 v} f_{\etae x_1 x}^{\phantom{\etae} \K_1 \K} {Q_Y}_{n x \K}^{n' v l_1}(z) \Big)
	\nn
	&\phantom{{\bar{\beta}_{x_1 \K_1}^{v_1 l_1}(z)}^{III}=\,} + \frac{1}{V} \sum_{x \K} \Big( J_0^\t{e} d |u_0^\t{e}(z)|^2 f_{-\etah x_1 x}^{\phantom{-\etah} \K_1 \K}
	+ J_0^\t{h} d |u_0^\t{h}(z)|^2 f_{\etae x_1 x}^{\phantom{\etae} \K_1 \K} \Big) \bar{Y}_{x \K}^{v_1 l_1}(z),
	\an
&{b_{n_1 x_1 \K_1 x_2 \K_2}^{n_2 l_1 v_1 v_2 l_2}(z)}^{I} 
	= \Jsd d |u_0^\t{e}(z)|^2 \sum_{l n} f_{-\etah x_2 x_1}^{\phantom{-\etah} \K_2 \K_1} \Big( 
	\mbf S_{n_2 n} M_{n_1}^{n}(z) \cdot \mbf s^\t{e}_{l_2 l} N_{x_1 \K_1}^{l_1 v_1 v_2 l}
	- \mbf S_{n n_1} M_{n}^{n_2}(z) \cdot \mbf s^\t{e}_{l l_1} N_{x_2 \K_2}^{l v_1 v_2 l_2} \Big)
	\nn
	&\phantom{{b_{n_1 x_1 \K_1 x_2 \K_2}^{n_2 l_1 v_1 v_2 l_2}(z)}^{I}=\,} + \Jpd d |u_0^\t{h}(z)|^2 \sum_{v n} f_{\etae x_2 x_1}^{\phantom{\etae} \K_2 \K_1} \Big( 
	\mbf S_{n_2 n} M_{n_1}^{n}(z) \cdot \mbf s^\t{h}_{v_2 v} N_{x_1 \K_1}^{l_1 v_1 v l_2}
	- \mbf S_{n n_1} M_{n}^{n_2}(z) \cdot \mbf s^\t{h}_{v v_1} N_{x_2 \K_2}^{l_1 v v_2 l_2} \Big)
	\nn
	&\phantom{{b_{n_1 x_1 \K_1 x_2 \K_2}^{n_2 l_1 v_1 v_2 l_2}(z)}^{I}=\,} 
	+ M_{n_1}^{n_2}(z) \Big( J_0^\t{e} d|u_0^\t{e}(z)|^2 f_{-\etah x_2 x_1}^{\phantom{-\etah} \K_2 \K_1} 
	+ J_0^\t{h} d|u_0^\t{h}(z)|^2 f_{\etae x_2 x_1}^{\phantom{\etae} \K_2 \K_1} \Big) 
	\Big( N_{x_1 \K_1}^{l_1 v_1 v_2 l_2} - N_{x_2 \K_2}^{l_1 v_1 v_2 l_2} \Big),
	\an
&{b_{n_1 x_1 \K_1 x_2 \K_2}^{n_2 l_1 v_1 v_2 l_2}(z)}^{II} 
	= \sum_{l} \hbar\bs\ome \cdot \Big( \mbf s^\t{e}_{l_2 l} {Q_N}_{n_1 x_1 \K_1 x_2 \K_2}^{n_2 l_1 v_1 v_2 l}(z)
	- \mbf s^\t{e}_{l l_1} {Q_N}_{n_1 x_1 \K_1 x_2 \K_2}^{n_2 l v_1 v_2 l_2}(z) \Big)
	\nn
	&\phantom{{b_{n_1 x_1 \K_1 x_2 \K_2}^{n_2 l_1 v_1 v_2 l_2}(z)}^{II}=\,} 
	+ \sum_{v} \hbar\bs\omh \cdot \Big( \mbf s^\t{h}_{v_2 v} {Q_N}_{n_1 x_1 \K_1 x_2 \K_2}^{n_2 l_1 v_1 v l_2}(z) 
	- \mbf s^\t{h}_{v v_1} {Q_N}_{n_1 x_1 \K_1 x_2 \K_2}^{n_2 l_1 v v_2 l_2}(z) \Big)
	\nn
	&\phantom{{b_{n_1 x_1 \K_1 x_2 \K_2}^{n_2 l_1 v_1 v_2 l_2}(z)}^{II}=\,} 
	+ \sum_{n} \hbar\bs\omMn \cdot \Big( \mbf S_{n_2 n} {Q_N}_{n_1 x_1 \K_1 x_2 \K_2}^{n l_1 v_1 v_2 l_2}(z)
	- \mbf S_{n n_1} {Q_N}_{n x_1 \K_1 x_2 \K_2}^{n_2 l_1 v_1 v_2 l_2}(z) \Big),
	\an
&{b_{n_1 x_1 \K_1 x_2 \K_2}^{n_2 l_1 v_1 v_2 l_2}(z)}^{III} 
	= \frac{\Jsd}{V} d |u_0^\t{e}(z)|^2 \sum_{\substack{l n \\ x \K}} \Big( 
	\mbf S_{n_2 n} \cdot \mbf s^\t{e}_{l_2 l} f_{-\etah x_2 x}^{\phantom{-\etah} \K_2 \K} {Q_N}_{n_1 x_1 \K_1 x \K}^{n l_1 v_1 v_2 l}(z)
	- \mbf S_{n n_1} \cdot \mbf s^\t{e}_{l l_1} f_{-\etah x x_1}^{\phantom{-\etah} \K \K_1} {Q_N}_{n x \K x_2 \K_2}^{n_2 l v_1 v_2 l_2}(z) \Big)
	\nn
	&\phantom{{b_{n_1 x_1 \K_1 x_2 \K_2}^{n_2 l_1 v_1 v_2 l_2}(z)}^{III}=\,} 
	+ \frac{\Jpd}{V} d |u_0^\t{h}(z)|^2 \sum_{\substack{v n \\ x \K}} \Big( 
	\mbf S_{n_2 n} \cdot \mbf s^\t{h}_{v_3 v} f_{\etae x_2 x}^{(-\etae) \K_2 \K} {Q_N}_{n_1 x_1 \K_1 x \K}^{n l_1 v_1 v l_2}(z)
	- \mbf S_{n n_1} \cdot \mbf s^\t{h}_{v v_2} f_{\etae x x_1}^{\phantom{\etae} \K \K_1} {Q_N}_{n x \K x_2 \K_2}^{n_2 l_1 v v_2 l_2}(z) \Big)
	\nn
	&\phantom{{b_{n_1 x_1 \K_1 x_2 \K_2}^{n_2 l_1 v_1 v_2 l_2}(z)}^{III}=\,} 
	+ \frac{J_0^\t{e}}{V} \sum_{x \K} d|u_0^\t{e}(z)|^2 
	\Big( f_{-\etah x_2 x}^{\phantom{-\etah} \K_2 \K} {Q_N}_{n_1 x_1 \K_1 x \K}^{n_2 l_1 v_1 v_2 l_2}(z) 
	- f_{-\etah x x_1}^{\phantom{-\etah} \K \K_1} {Q_N}_{n_1 x \K x_2 \K_2}^{n_2 l_1 v_1 v_2 l_2}(z) \Big)
	\nn
	&\phantom{{b_{n_1 x_1 \K_1 x_2 \K_2}^{n_2 l_1 v_1 v_2 l_2}(z)}^{III}=\,} 
	+ \frac{J_0^\t{h}}{V} \sum_{x \K} d|u_0^\t{h}(z)|^2 
	\Big( f_{\etae x_2 x}^{\phantom{\etae} \K_2 \K} {Q_N}_{n_1 x_1 \K_1 x \K}^{n_2 l_1 v_1 v_2 l_2}(z) 
	- f_{\etae x x_1}^{\phantom{\etae} \K \K_1} {Q_N}_{n_1 x \K x_2 \K_2}^{n_2 l_1 v_1 v_2 l_2}(z) \Big),
	\an
&{\bar{b}_{x_1 \K_1 x_2 \K_2}^{l_1 v_1 v_2 l_2}(z)}^{I}
	= \frac{\Jsd \NMn}{V} d |u_0^\t{e}(z)|^2 \sum_{l n n'} \mbf S_{n n'} M_{n}^{n'}(z) \cdot f_{-\etah x_2 x_1}^{\phantom{-\etah} \K_2 \K_1} 
	\Big( \mbf s^\t{e}_{l_2 l} N_{x_1 \K_1}^{l_1 v_1 v_2 l} - \mbf s^\t{e}_{l l_1} N_{x_2 \K_2}^{l v_1 v_2 l_2} \Big)
	\nn
	&\phantom{{\bar{b}_{x_1 \K_1 x_2 \K_2}^{l_1 v_1 v_2 l_2}(z)}^{I}=\,} 
	+ \frac{\Jpd \NMn}{V} d |u_0^\t{h}(z)|^2 \sum_{v n n'} \mbf S_{n n'} M_{n}^{n'}(z) \cdot 
	f_{\etae x_2 x_1}^{\phantom{\etae} \K_2 \K_1} \Big( \mbf s^\t{h}_{v_2 v} N_{x_1 \K_1}^{l_1 v_1 v l_2} - \mbf s^\t{h}_{v v_1} N_{x_2 \K_2}^{l_1 v v_2 l_2} \Big)
	\nn
	&\phantom{{\bar{b}_{x_1 \K_1 x_2 \K_2}^{l_1 v_1 v_2 l_2}(z)}^{I}=\,} 
	+ \Big( J_0^\t{e} d|u_0^\t{e}(z)|^2 f_{-\etah x_2 x_1}^{\phantom{-\etah} \K_2 \K_1} 
	+ J_0^\t{h} d|u_0^\t{h}(z)|^2 f_{\etae x_2 x_1}^{\phantom{\etae} \K_2 \K_1} \Big) 
	\Big( N_{x_1 \K_1}^{l_1 v_1 v_2 l_2} - N_{x_2 \K_2}^{l_1 v_1 v_2 l_2} \Big),
	\an
&{\bar{b}_{x_1 \K_1 x_2 \K_2}^{l_1 v_1 v_2 l_2}(z)}^{II}
	= \sum_{l} \hbar\bs\ome \cdot \Big( \mbf s^\t{e}_{l_2 l} \bar{N}_{x_1 \K_1 x_2 \K_2}^{l_1 v_1 v_2 l}(z) - \mbf s^\t{e}_{l l_1} \bar{N}_{x_1 \K_1 x_2 \K_2}^{l v_1 v_2 l_2}(z) \Big)
	\nn
	&\phantom{{\bar{b}_{x_1 \K_1 x_2 \K_2}^{l_1 v_1 v_2 l_2}(z)}^{II}=\,} 
	+ \sum_{v} \hbar\bs\omh \cdot \Big( \mbf s^\t{h}_{v_2 v} \bar{N}_{x_1 \K_1 x_2 \K_2}^{l_1 v_1 v l_2}(z) - \mbf s^\t{h}_{v v_1} \bar{N}_{x_1 \K_1 x_2 \K_2}^{l_1 v v_2 l_2}(z) \Big),
	\an
&{\bar{b}_{x_1 \K_1 x_2 \K_2}^{l_1 v_1 v_2 l_2}(z)}^{III}
	= \frac{\Jsd \NMn}{V^2} d |u_0^\t{e}(z)|^2 \sum_{\substack{l n n' \\ x \K}} \mbf S_{n n'} \cdot 
	\Big( \mbf s^\t{e}_{l_2 l} f_{-\etah x_2 x}^{\phantom{-\etah} \K_2 \K} {Q_N}_{n_1 x_1 \K_1 x \K}^{n l_1 v_1 v_2 l}(z) 
	- \mbf s^\t{e}_{l l_1} f_{-\etah x x_1}^{\phantom{-\etah} \K \K_1} {Q_N}_{n x \K x_2 \K_2}^{n_2 l v_1 v_2 l_2}(z) \Big)
	\nn
	&\phantom{{\bar{b}_{x_1 \K_1 x_2 \K_2}^{l_1 v_1 v_2 l_2}(z)}^{III}=\,} 
	+ \frac{\Jpd \NMn}{V^2} d |u_0^\t{h}(z)|^2 \sum_{\substack{v n n' \\ x \K}} \mbf S_{n n'} \cdot 
	\Big( \mbf s^\t{h}_{v_2 v} f_{\etae x_2 x}^{\phantom{\etae} \K_2 \K} {Q_N}_{n_1 x_1 \K_1 x \K}^{n l_1 v_1 v l_2}(z)
	- \mbf s^\t{h}_{v v_1} f_{\etae x x_1}^{\phantom{\etae} \K \K_1} {Q_N}_{n x \K x_2 \K_2}^{n_2 l_1 v v_2 l_2}(z) \Big)
	\nn
	&\phantom{{\bar{b}_{x_1 \K_1 x_2 \K_2}^{l_1 v_1 v_2 l_2}(z)}^{III}=\,} 
	+ \frac{J_0^\t{e}}{V} d|u_0^\t{e}(z)|^2 \sum_{x \K} 
	\Big( f_{-\etah x_2 x}^{\phantom{-\etah} \K_2 \K} \bar{N}_{x_1 \K_1 x \K}^{l_1 v_1 v_2 l_2}(z) 
	- f_{-\etah x x_1}^{\phantom{-\etah} \K \K_1} \bar{N}_{x \K x_2 \K_2}^{l_1 v_1 v_2 l_2}(z) \Big)
	\nn
	&\phantom{{\bar{b}_{x_1 \K_1 x_2 \K_2}^{l_1 v_1 v_2 l_2}(z)}^{III}=\,} 
	+ \frac{J_0^\t{h}}{V} d|u_0^\t{h}(z)|^2 \sum_{x \K}
	\Big( f_{\etae x_2 x}^{\phantom{\etae} \K_2 \K} \bar{N}_{x_1 \K_1 x \K}^{l_1 v_1 v_2 l_2}(z) 
	- f_{\etae x x_1}^{\phantom{\etae} \K \K_1} \bar{N}_{x \K x_2 \K_2}^{l_1 v_1 v_2 l_2}(z) \Big).
\end{align}
\end{subequations}
\end{widetext}

\newpage
\section{Quantum kinetic equations of motion with pinned hole spin}
\label{ap Quantum kinetic equations of motion with pinned hole spin}

In this section, we provide the equations of motion corresponding to the variables defined in Eqs.~\eqref{eq dynamical variables summed} after performing an angle-averaging
in $\K$ space.
Using the Einstein summation convention, the equations read:

\begin{widetext}
\begin{subequations}
\label{eq EOM for summed variables}
\begin{align}
\label{eq EOM for summed variables n}
\ddt n_{x_1 K_1} &= 
	\frac{1}{\hbar}\mbf E \cdot \mbf M 2\Im\big[y_{x_1}^\uparrow \phi_{x_1} \big] \delta_{K_1,0}
	- \frac{\Jsd \NMn}{\hbar V^2} \sum_{x K} 2\Im\big[ Q_{-\etah i x K}^{\phantom{-\etah} i x_1 K_1} \big]
	+ \frac{\Jpd \NMn}{\hbar V^2} \sum_{x K} \Im\big[ Q_{\etae z x K}^{\phantom{\etae} 0 x_1 K_1} \big]
	\nn
	&\phantom{=\;} - \frac{J_0^\t{e} \NMn}{\hbar V^2} \sum_{x K} 2\Im\big[ Z_{-\etah \phantom{0} x K}^{\phantom{-\etah} 0 x_1 K_1} \big]
	- \frac{J_0^\t{h} \NMn}{\hbar V^2} \sum_{x K} 2\Im\big[ Z_{\etae \phantom{0} x K}^{\phantom{\etae} 0 x_1 K_1} \big],
	\an
\label{eq EOM for summed variables s}
\ddt s_{x_1 K_1}^l &= 
	\frac{1}{\hbar}\mbf E \cdot \mbf M \Big( \Im\big[ y_{x_1}^\uparrow \phi_{x_1} \big] \delta_{K_1,0}\delta_{l,z}
	+ \Im\big[ y_{x_1}^\downarrow \phi_{x_1} \big] \delta_{K_1,0}\delta_{l,x}
	- \Re\big[ y_{x_1}^\downarrow \phi_{x_1} \big] \delta_{K_1,0}\delta_{l,y} \Big)
	+ \epsilon_{ijl} \ome^i s_{x_1 K_1}^j
	\nn
	&\phantom{=\;} + \frac{\Jsd \NMn}{\hbar V^2} \sum_{x K} \Big( \epsilon_{ijl} \Re\big[Q_{-\etah i x K}^{\phantom{-\etah} j x_1 K_1}\big] 
	- \frac{1}{2} \Im\big[Q_{-\etah l x K}^{\phantom{-\etah} 0 x_1 K_1}\big] \Big)
	+ \frac{\Jpd \NMn}{\hbar V^2} \sum_{x K} \Im\big[ Q_{\etae z x K}^{\phantom{\etae} l x_1 K_1} \big] 
	\nn
	&\phantom{=\;} - \frac{J_0^\t{e} \NMn}{\hbar V^2} \sum_{x K} 2\Im\big[ Z_{-\etah \phantom{l} x K}^{\phantom{-\etah} l x_1 K_1} \big]
	- \frac{J_0^\t{h} \NMn}{\hbar V^2} \sum_{x K} 2\Im\big[ Z_{\etae \phantom{l} x K}^{\phantom{\etae} l x_1 K_1} \big],
	\an
\ddt y_{x_1}^\ud &= 
	\frac{i}{\hbar}\mbf E \cdot \mbf M \phi_{x_1} \delta_{\ud, \uparrow}
	-i \Big( \omega_{0 x_1} \pm \frac{1}{2} \ome^z - \frac{1}{2} \omh^z + \frac{(J_0^\t{e}+J_0^\t{h}) \NMn}{\hbar V} \Big) y_{x_1}^\ud
	-i \frac{1}{2} \ome^\mp y_{x_1}^\du
	\nn
	&\phantom{=\;} - i \frac{\Jsd \NMn}{2\hbar V^2} \sum_{x K} \Big(\pm q_{-\etah z x K}^{\phantom{-\etah} \ud x_1} + q_{-\etah \mp x K}^{\phantom{-\etah} \du x_1}\Big)
	+ i \frac{\Jpd \NMn}{2\hbar V^2} \sum_{x K} q_{\etae z x K}^{\phantom{\etae} \ud x_1}
	\nn
	&\phantom{=\;} - i \frac{J_0^\t{e} \NMn}{\hbar V^2} \sum_{x K} z_{-\etah x K}^{\phantom{-\etah} \ud x_1}
	- i \frac{J_0^\t{h} \NMn}{\hbar V^2} \sum_{x K} z_{\etae x K}^{\phantom{\etae} \ud x_1},
	\an
\ddt q_{\eta l x_1 K_1}^{\phantom{\eta} \ud x_2} &= 
	- i \Big( \omega_{x_1 K_1} \pm \frac{1}{2} \ome^z - \frac{1}{2} \omh^z + \frac{I (J_0^\t{e}+J_0^\t{h}) \NMn}{\hbar V} \Big) q_{\eta l x_1 K_1}^{\phantom{\eta} \ud x_2}
	-i \frac{1}{2} \ome^\mp q_{\eta l x_1 K_1}^{\phantom{\eta} \du x_2}
	+ \epsilon_{ijl} \omMn^i q_{\eta j x_1 K_1}^{\phantom{\eta} \ud x_2}
	\nn
	&\phantom{=\;} - i \frac{I \Jsd}{2\hbar} F_{\phantom{-}\eta\phantom{_h} x_2 x_1}^{-\etah 0 K_1} \Big(\pm\langle S^lS^z \rangle y_{x_2}^\ud + \langle S^lS^\mp \rangle y_{x_2}^\du\Big)
	+ i \frac{I \Jpd}{2\hbar} \langle S^lS^z \rangle F_{\eta\phantom{_e} x_2 x_1}^{\etae 0 K_1} y_{x_2}^\ud
	\nn
	&\phantom{=\;} - i \frac{I}{\hbar} \langle S^l \rangle \Big( J_0^\t{e} F_{\phantom{-}\eta\phantom{_h} x_2 x_1}^{-\etah 0 K_1} + J_0^\t{h} F_{\eta\phantom{_e} x_2 x_1}^{\etae 0 K_1} \Big) 
	y_{x_2}^\ud,
	\an
\ddt z_{\eta x_1 K_1}^{\phantom{\eta} \ud x_2} &= 
	- i \Big( \omega_{x_1 K_1} \pm \frac{1}{2} \ome^z - \frac{1}{2} \omh^z + \frac{I (J_0^\t{e}+J_0^\t{h}) \NMn}{\hbar V} \Big) z_{\eta x_1 K_1}^{\phantom{\eta} \ud x_2}
	- i\frac{1}{2} \ome^\mp z_{\eta x_1 K_1}^{\phantom{\eta} \du x_2}
	\nn
	&\phantom{=\;} - i \frac{I \Jsd}{2\hbar} F_{\phantom{-}\eta\phantom{_h} x_2 x_1}^{-\etah 0 K_1} \Big(\pm\langle S^z \rangle y_{x_2}^\ud + \langle S^\mp \rangle y_{x_2}^\du\Big)
	+ i \frac{I \Jpd}{2\hbar} \langle S^z\rangle F_{\eta\phantom{_e} x_2 x_1}^{\etae 0 K_1} y_{x_2}^\ud
	\nn
	&\phantom{=\;} - i \frac{I}{\hbar} \Big( J_0^\t{e} F_{\phantom{-}\eta\phantom{_h} x_2 x_1}^{-\etah 0 K_1} + J_0^\t{h} F_{\eta\phantom{_e} x_2 x_1}^{\etae 0 K_1} \Big) 
	y_{x_2}^\ud,
	\an
\label{eq EOM for summed variables Q1}
\ddt Q_{\eta l x_1 K_1}^{\phantom{\eta} 0 x_2 K_2} &= 
	- i \big(\omega_{x_2 K_2} - \omega_{x_1 K_1}\big) Q_{\eta l x_1 K_1}^{\phantom{\eta} 0 x_2 K_2}
	+ \epsilon_{ijl} \omMn^i Q_{\eta j x_1 K_1}^{\phantom{\eta} 0 x_2 K_2}
	+ \frac{i}{2\hbar}\mbf E \cdot \mbf M \Big(\big(q_{\eta l x_1 K_1}^{\phantom{\eta} \uparrow x_2} \phi_{x_2}\big)^* \delta_{K_2,0} 
	- q_{\eta l x_2 K_2}^{\phantom{\eta} \uparrow x_1} \phi_{x_1}\delta_{K_1,0}\Big)
	\nn
	&\phantom{=\;}+ i \frac{I \Jsd}{\hbar} F_{\phantom{-}\eta\phantom{_h} x_1 x_2}^{-\etah K_1 K_2} \Big( \langle S^iS^l \rangle s_{x_2 K_2}^i - \langle S^lS^i \rangle s_{x_1 K_1}^i \Big)
	- i \frac{I \Jpd}{\hbar} F_{\eta\phantom{_e} x_1 x_2}^{\etae K_1 K_2} \frac{1}{2} \Big( \langle S^zS^l \rangle n_{x_2 K_2} - \langle S^lS^z \rangle n_{x_1 K_1} \Big)
	\nn
	&\phantom{=\;}+ i \frac{I}{\hbar} \langle S^l \rangle \Big( J_0^\t{e} F_{\phantom{-}\eta\phantom{_h} x_1 x_2}^{-\etah K_1 K_2} 
	+ J_0^\t{h} F_{\eta\phantom{_e} x_1 x_2}^{\etae K_1 K_2} \Big) \big( n_{x_2 K_2} - n_{x_1 K_1} \big),
	\an
\label{eq EOM for summed variables Q2}
\ddt Q_{\eta l x_1 K_1}^{\phantom{\eta} m x_2 K_2} &= 
	- i \big(\omega_{x_2 K_2} - \omega_{x_1 K_1}\big) Q_{\eta l x_1 K_1}^{\phantom{\eta} m x_2 K_2}
	+ \epsilon_{ijm} \ome^i Q_{\eta l x_1 K_1}^{\phantom{\eta} j x_2 K_2}
	+ \epsilon_{ijl} \omMn^i Q_{\eta j x_1 K_1}^{\phantom{\eta} m x_2 K_2}
	\nn
	&\phantom{=\;} + \frac{i}{2\hbar}\mbf E \cdot \mbf M \bigg[ \Big(\big(q_{\eta l x_1 K_1}^{\phantom{\eta} \uparrow x_2} \phi_{x_2}\big)^* \delta_{K_2,0} 
	\! - \! q_{\eta l x_2 K_2}^{\phantom{\eta} \uparrow x_1} \phi_{x_1}\delta_{K_1,0}\Big) \delta_{m,z}
	+ \Big(\big(q_{\eta l x_1 K_1}^{\phantom{\eta} \downarrow x_2} \phi_{x_2}\big)^* \delta_{K_2,0} 
	\! - \! q_{\eta l x_2 K_2}^{\phantom{\eta} \downarrow x_1} \phi_{x_1}\delta_{K_1,0}\Big) \delta_{m,x}
	\nn
	 &\phantom{=\;} + i \Big(\big(q_{\eta l x_1 K_1}^{\phantom{\eta} \downarrow x_2} \phi_{x_2}\big)^* \delta_{K_2,0} 
	\! + \! q_{\eta l x_2 K_2}^{\phantom{\eta} \downarrow x_1} \phi_{x_1}\delta_{K_1,0}\Big) \delta_{m,y} \bigg]
	- i \frac{I \Jpd}{\hbar} F_{\eta\phantom{_e} x_1 x_2}^{\etae K_1 K_2} \frac{1}{2} \Big( \langle S^zS^l \rangle s_{x_2 K_2}^m \! - \! \langle S^lS^z \rangle s_{x_1 K_1}^m \Big)
	\nn
	&\phantom{=\;}+ i \frac{I \Jsd}{2 \hbar} F_{\phantom{-}\eta\phantom{_h} x_1 x_2}^{-\etah K_1 K_2} \Big( 
	\langle S^iS^l \rangle \big( \frac{1}{2}\delta_{i,m} n_{x_2 K_2} - i \epsilon_{ijm} s_{x_2 K_2}^j \big) 
	- \langle S^lS^i \rangle \big( \frac{1}{2}\delta_{i,m} n_{x_1 K_1} + i \epsilon_{ijm} s_{x_1 K_1}^j \big)\Big)
	\nn
	&\phantom{=\;}+ i \frac{I}{\hbar} \langle S^l \rangle \Big( J_0^\t{e} F_{\phantom{-}\eta\phantom{_h} x_1 x_2}^{-\etah K_1 K_2} 
	+ J_0^\t{h} F_{\eta\phantom{_e} x_1 x_2}^{\etae K_1 K_2} \Big) \big( s_{x_2 K_2}^m - s_{x_1 K_1}^m \big),
	\an
\label{eq EOM for summed variables Z1}
\ddt Z_{\eta \phantom{0} x_1 K_1}^{\phantom{\eta} 0 x_2 K_2} &=
	- i \big(\omega_{x_2 K_2} - \omega_{x_1 K_1}\big) Z_{\eta \phantom{0} x_1 K_1}^{\phantom{\eta} 0 x_2 K_2}
	+ \frac{i}{2\hbar}\mbf E \cdot \mbf M \Big(\big(z_{\eta x_1 K_1}^{\phantom{\eta} \uparrow x_2} \phi_{x_2}\big)^* \delta_{K_2,0} 
	- z_{\eta x_2 K_2}^{\phantom{\eta} \uparrow x_1} \phi_{x_1}\delta_{K_1,0}\Big)
	\nn
	&\phantom{=\;}+ i \frac{I \Jsd \NMn}{\hbar V} F_{\phantom{-}\eta\phantom{_h} x_1 x_2}^{-\etah K_1 K_2} \langle S^i \rangle \big( s_{x_2 K_2}^i - s_{x_1 K_1}^i \big)
	- i \frac{I \Jpd \NMn}{\hbar V} F_{\eta\phantom{_e} x_1 x_2}^{\etae K_1 K_2} \frac{1}{2} \langle S^z \rangle \big( n_{x_2 K_2} - n_{x_1 K_1} \big)
	\nn
	&\phantom{=\;}+ i \frac{I}{\hbar} \Big( J_0^\t{e} F_{\phantom{-}\eta\phantom{_h} x_1 x_2}^{-\etah K_1 K_2} 
	+ J_0^\t{h} F_{\eta\phantom{_e} x_1 x_2}^{\etae K_1 K_2} \Big) \big( n_{x_2 K_2} - n_{x_1 K_1} \big),
	\an
\label{eq EOM for summed variables Z2}
\ddt Z_{\eta \phantom{l} x_1 K_1}^{\phantom{\eta} l x_2 K_2} &=
	- i \big(\omega_{x_2 K_2} - \omega_{x_1 K_1}\big) Z_{\eta \phantom{l} x_1 K_1}^{\phantom{\eta} l x_2 K_2}
	+ \epsilon_{ijl} \ome^i Z_{\eta \phantom{j} x_1 K_1}^{\phantom{\eta} j x_2 K_2}
	+ \frac{i}{2\hbar}\mbf E \cdot \mbf M \bigg[ \Big(\big(z_{\eta x_1 K_1}^{\phantom{\eta} \uparrow x_2} \phi_{x_2}\big)^* \delta_{K_2,0} 
	- z_{\eta x_2 K_2}^{\phantom{\eta} \uparrow x_1} \phi_{x_1}\delta_{K_1,0}\Big)\delta_{l,z}
	\nn
	&\phantom{=\;} + \Big(\big(z_{\eta x_1 K_1}^{\phantom{\eta} \downarrow x_2} \phi_{x_2}\big)^* \delta_{K_2,0} 
	- z_{\eta x_2 K_2}^{\phantom{\eta} \downarrow x_1} \phi_{x_1}\delta_{K_1,0}\Big)\delta_{l,x}
	+ i \Big(\big(z_{\eta x_1 K_1}^{\phantom{\eta} \downarrow x_2} \phi_{x_2}\big)^* \delta_{K_2,0} 
	+ z_{\eta x_2 K_2}^{\phantom{\eta} \downarrow x_1} \phi_{x_1}\delta_{K_1,0}\Big)\delta_{l,y} \bigg]
	\nn
	&\phantom{=\;}+ i \frac{I \Jsd}{2 \hbar} F_{\phantom{-}\eta\phantom{_h} x_1 x_2}^{-\etah K_1 K_2} \langle S^i \rangle \Big( 
	\big( \frac{1}{2}\delta_{i,l} n_{x_2 K_2} - i \epsilon_{ijl} s_{x_2 K_2}^j \big) 
	- \big( \frac{1}{2}\delta_{i,l} n_{x_1 K_1} + i \epsilon_{ijl} s_{x_1 K_1}^j \big)\Big)
	\nn
	&\phantom{=\;}- i \frac{I \Jpd}{\hbar} F_{\eta\phantom{_e} x_1 x_2}^{\etae K_1 K_2} \frac{1}{2} \langle S^z \rangle \big( s_{x_2 K_2}^l - s_{x_1 K_1}^l \big)
	+ i \frac{I}{\hbar} \Big( J_0^\t{e} F_{\phantom{-}\eta\phantom{_h} x_1 x_2}^{-\etah K_1 K_2}
	+ J_0^\t{h} F_{\eta\phantom{_e} x_1 x_2}^{\etae K_1 K_2} \Big) \big( s_{x_2 K_2}^l - s_{x_1 K_1}^l \big).
\end{align}
\end{subequations}
\end{widetext}

%%%%%%%%%%%%%%%%%%%%%%%%%%%%%%%%%%%%%%%%%%%%%%%%%%
% references and end of document
%%%%%%%%%%%%%%%%%%%%%%%%%%%%%%%%%%%%%%%%%%%%%%%%%%
\bibliography{references.bib}
\end{document}